\documentclass[11pt,a4paper]{article}
\usepackage[margin=1in]{geometry}
\usepackage[utf8]{inputenc}
\usepackage[T1]{fontenc}
\usepackage{lmodern}
\usepackage{amsmath,amssymb}
\usepackage{xcolor}
\usepackage{graphicx}
\usepackage{cite}
\usepackage{caption}
\usepackage{doi}
\usepackage{hyperref}
\hypersetup{colorlinks, linkcolor={red!50!black}, citecolor={blue!50!black}, urlcolor={blue!80!black}}
\usepackage{bm}
\usepackage{siunitx}
\usepackage{array}
\usepackage{orcidlink}
\ifdefined\locked\else\newenvironment{locked}{}{}\fi
\providecommand{\lockmark}[1]{}
\begin{document}
\title{Edge quantum weight of finite Chern flakes and its topological floor}
\author{Rohit Narula\,\orcidlink{0000-0001-5304-4873}\\[0.5ex]
{\normalsize Department of Physics, Indian Institute of Technology Delhi, New Delhi, India}\\
{\normalsize\href{mailto:rnarula@physics.iitd.ac.in}{rnarula@physics.iitd.ac.in}}}
\date{}
\maketitle
\begin{abstract}
The zero-point energy of an $LC$ circuit measures the polarization fluctuation of an insulator coupled to it. In a crystal this fluctuation is the bulk quantum weight $K_\mathrm{bulk}$, which the Chern number $C$ bounds below by $|C|$. We show that a finite Chern flake adds an edge term that survives as the flake grows. The low-frequency transitions of its chiral edge have squared dipole matrix elements that sum to an edge weight $K_\mathrm{edge}\geq|C|$ per unit area, a floor that follows from the positive absorption of both circular polarizations and the known cancellation of circular dichroism between edge and bulk. At weak coupling, which a growing flake eventually leaves, a circuit slower than the edge modes reads $K_\mathrm{bulk}+K_\mathrm{edge}$, averaged over two field directions, and a faster one approaches $K_\mathrm{bulk}$, so the reading depends on whether the flake is made large or the circuit slow first. In exact free-fermion calculations on lattice flakes with $|C|=1$ and 2 the summed circular dichroism of the edge sector extrapolates to $|C|$ in magnitude, so that the floor is $|C|$. Above the floor, each edge channel adds a number fixed by the boundary shape and its edge velocities, which square and rectangular lattice flakes with one channel follow to within 0.003, while the circular dichroism counts only the net chirality. The avoided crossing of the circuit with the edge mode gives the edge weight in units of the Hall conductance quantum, up to a geometric factor of the capacitor.
\end{abstract}

\section{Introduction}

The electric polarization of an insulator fluctuates in its ground state. The size of this fluctuation is a geometric property of the ground state known as the quantum weight $K$~\cite{onishi2026vacuum}, and the Chern number $C$ of a filled band bounds it from below~\cite{onishi2024bound}. Onishi and Fu have proposed to weigh it without a probe field. The zero-point field of a superconducting $LC$ circuit gives an insulator coupled to it a zero-point energy proportional to that fluctuation, which appears as a force between the circuit and the material~\cite{onishi2026vacuum}. The Chern insulators at hand are finite flakes, and a Chern flake has a chiral edge whose excitations lie far below the bulk gap. This paper asks what such a circuit reads on a finite Chern flake. The edge turns out to add a term of its own to the reading, a term that at weak coupling does not shrink as the flake grows and tends to at least $|C|$ per unit area. The circuit sheds it as it becomes faster than the edge, and its size above $|C|$ is fixed by the shape of the flake and the velocities of its edge channels.

An insulator is distinguished from a metal by the localization of its many-electron ground state~\cite{kohn1964theory,resta1999localization}, the size of which is set by the quantum fluctuation of the electric polarization~\cite{souza2000polarization}. From the second cumulant of the polarization one defines a localization tensor, which is the integrated quantum metric of the occupied states~\cite{souza2000polarization} and the gauge-invariant part of the spread of their Wannier functions~\cite{marzari1997maximally}. That metric is the Riemannian metric which the overlap of neighbouring states defines on the manifold of quantum states~\cite{provost1980riemannian}. Onishi and Fu recast this quantity as the quantum weight $K$, the coefficient of the quadratic term of the static structure factor at long wavelength~\cite{onishi2025weight}, and showed that it is a natural generalization of the position fluctuation to extended systems~\cite{onishi2025weight}. The same geometry sets an intrinsic capacitance through the ratio of the quantum metric to the gap~\cite{komissarov2024capacitance}, and it enters the instantaneous dielectric response~\cite{verma2025instantaneous} and resonant optical responses~\cite{ahn2022riemannian}. The long-wavelength scaling of the structure factor classifies ground-state phases~\cite{hyperuniformity2025}, and its fourth-order term is bounded as well~\cite{geombound2025}. Related bounds hold for every linear response and bound the Drude weight by the orbital magnetization~\cite{qgbounds2025observables}, give a lower bound on the electric susceptibility of Chern insulators~\cite{magbounds2026}, and extend dichroic sum rules to the orbital and heat magnetization~\cite{chiralthermo2026dichroism}. These developments make the quantum weight a static, measurable property of an insulating ground state~\cite{torma2025review}, and spectroscopic responses can read it in synthetic quantum systems as well~\cite{ozawa2019probing}. Inelastic x-ray scattering has measured the quantum weight of a bulk insulator, LiF~\cite{balut2025quantum}. Real-space markers map quantum-geometric responses such as the superfluid weight to individual lattice sites~\cite{porlles2025quantum}, and band geometry has been formulated in terms of gauge-invariant projectors~\cite{mitscherling2025gauge}.

Topology places a floor under this geometry. For a filled band, an integer, the Chern number $C$, quantizes the Hall conductance~\cite{thouless1982quantized}, and a lattice model need not host Landau levels for the realization of this quantization~\cite{haldane1988model}. The trace of the quantum metric bounds the Berry curvature pointwise~\cite{roy2014band}, so for a Chern band $|C|$ bounds the integrated quantum metric from below~\cite{peotta2015superfluidity,ozawa2021relations}, the quantum weight obeys this bound~\cite{onishi2024bound}, and so does the long-wavelength structure factor of interacting systems~\cite{onishi2024structure}. Subsequent work has extended the bound beyond symmetry-protected phases~\cite{onishi2025extending}. A system saturates the bound when, at every frequency, its absorption of circularly polarized light involves only one handedness~\cite{onishi2024bound}, a regime that isotropic ideal Chern bands, with holomorphic Bloch states, attain~\cite{elliptic2026dichroism}. The Chern insulator of twisted MoTe$_2$ comes close to saturation, with a nearly perfect magnetic circular dichroism of its excitonic absorption attributed to a nearly saturated bound~\cite{qiu2025quantum}.

Onishi and Fu proposed to weigh this quantity without any probe field~\cite{onishi2026vacuum}. In their theory the zero-point fluctuations of a superconducting $LC$ circuit induce in a coupled insulator an extensive zero-point energy proportional to the quantum fluctuation of its polarization. This produces a repulsive force between circuit and material and an attractive force on the capacitor plates~\cite{onishi2026vacuum}. This force is set by the many-body quantum metric of the electronic ground state, and for a gapped two-dimensional material that metric is identified with the quantum weight through the long-wavelength structure factor~\cite{onishi2026vacuum}. \begin{locked}Their theory considers ``a general electronic system with a finite energy gap $\Delta$ between the ground state $|\Psi_0\rangle$ and the first excited state,'' and the quantum-geometric force dominates ``in the adiabatic limit $\hbar\omega_p\ll\Delta$,'' where $\omega_p$ is the circuit frequency~\cite{onishi2026vacuum}.\end{locked} Natural targets are the zero-field Chern and fractional Chern insulators of twisted MoTe$_2$ and rhombohedral graphene~\cite{cai2023signatures,park2023observation,lu2024fractional}, which exist as finite flakes in van der Waals stacks. In such stacks the flakes and their gates form plasmonic cavities at gigahertz to terahertz frequencies, shaping the low-energy electrodynamics~\cite{kipp2025cavity}.

These samples are finite, and a finite Chern flake strains the identification of the circuit reading with the bulk quantum weight. Its lowest excitations are chiral edge modes rather than bulk excitations, and they can lie at the microwave frequencies of superconducting circuits. The zero-field edge magnetoplasmon of a magnetic topological insulator, a charge wave that circles the sample along its edge, has a fundamental mode near \qty{400}{\mega\hertz} for a disk \qty{330}{\micro\meter} across~\cite{mahoney2017zero}, with a dispersion set by the disk size~\cite{emp2024qah}, and a microwave-impedance signature is predicted at the sample boundary~\cite{mim2023chern}.

The sum rules known for these modes fix their antisymmetric, circular-dichroic dipole response, whereas the capacitor force reads the symmetric one, the response averaged over linear polarizations. The circular-dichroic absorption of a filled Chern band is quantized by $C$~\cite{tran2017heating,asteria2019measuring}, a fractional Chern insulator shows the fractional value~\cite{repellin2019detecting}, and moments of the dichroic spectrum are fixed by ground-state properties~\cite{souza2008dichroic,hallsumrule2025}. Sum rules for the off-diagonal conductivity of a solid in a magnetic field go back to Bennett and Stern~\cite{bennett1965faraday}. For a Chern insulator with boundaries the dichroic response of the edge states exactly cancels that of the bulk~\cite{tran2018quantized}. In any isolated, confined system the dichroic responses of edge and bulk cancel exactly, and the edge part, carried by the low-energy chiral modes, is quantized by the many-body Chern number~\cite{unal2024edge}. The edge and bulk collective modes of fractional Chern droplets can also be separated spectroscopically~\cite{fci2024spectroscopy}.

The symmetric response of a bounded Chern sample is not that of the bulk either. Marrazzo and Resta found that the localization tensor of a whole Chern flake still converges, but to a large value, because each boundary cell contributes a term proportional to the linear size of the flake and the number of such cells grows in the same way~\cite{marrazzo2019local}. For a quantum Hall droplet coupled through its dipole to a uniform microwave field, Cano and co-workers found a series of absorption peaks for each charged edge mode, spaced by the velocity of that mode, so that a droplet of known shape reveals the number and the velocities of its edge modes~\cite{cano2013microwave}. For anisotropic quantum Hall droplets the magnitude of each absorption peak has since been written as a Fourier mode, in the canonical angle of the confinement, of a function fixed by the shape of the droplet~\cite{oblak2024anisotropic}. Reference~\cite{onishi2026vacuum} nevertheless identifies the capacitor reading with the quantum weight, an intensive property of the bulk. Because the edge sum rule fixes only the dichroic response, it leaves open how much symmetric weight the edge adds to the zero-point force, what fixes that amount, and at what circuit frequency the reading returns to the bulk weight.

Here we show that the quantum weight which a zero-point circuit reads on a finite Chern flake depends on the order of two limits. An infinite flake at fixed circuit frequency gives the bulk weight $K_\mathrm{bulk}$. A slow circuit at fixed flake size, weakly coupled and slower than the edge magnetoplasmons, gives $K_\mathrm{bulk}+K_\mathrm{edge}$, averaged over two field directions, because the edge sector of the flake, its in-gap transitions, carries a symmetric dipole weight $K_\mathrm{edge}\geq|C|$ per unit area that survives as the flake grows. That the two limits differ is no surprise. The first excited state of a flake is an edge excitation, so the adiabatic condition of Ref.~\cite{onishi2026vacuum} is set for a flake by its edge and not by its bulk gap (Sec.~\ref{sec:theory}). What that condition leaves open is how much the edge adds, what fixes that amount, and at which circuit frequency it leaves the reading. The inequality is a corollary of known results. It follows from the edge--bulk cancellation of Ref.~\cite{unal2024edge} and the positivity of circular absorption at every frequency, which underlies the bulk bound~\cite{onishi2024bound}. Its area scaling follows from two known results, the dichroic edge response to a uniform circular drive, which is proportional to the area~\cite{unal2024edge}, and the extensive boundary term in the localization tensor of a Chern flake~\cite{marrazzo2019local}. What is new is the size of the edge term in the circuit reading and what fixes it. This comprises the frequency at which the term leaves the reading, the Dirichlet form of the edge weight above the floor with its closed values, the tests on lattice flakes, and a readout through the circuit frequency. An edge weight bounded below by $|C|$ enters the reading and drops out of it above the edge scale, the edge-magnetoplasmon frequency. At weak coupling and below the edge scale the edge weight enters the force with the same area scaling as the bulk weight, because the dipole moment of a perimeter mode grows with the linear size of the flake. In exact free-fermion calculations on flakes of an anisotropic Chern lattice model, resolved by frequency, the in-gap sector carries a Hall weight, a summed circular dichroism, that extrapolates to $-C$ and a symmetric weight that extrapolates to $|C|$ or above. For the flakes with the weaker of the two hopping anisotropies studied, the static circuit reading extrapolates to 2.27--2.32, nearly twice the torus bulk weight of 1.24, computed with periodic boundaries, and it falls onto the bulk curve, the reading of the periodic system, as the circuit frequency rises well above the edge scale. The reading within reach of a device is the circuit frequency and not the force. There the edge of a disk adds a capacitance equal to the Hall conductance divided by the edge frequency, times a geometric factor of the capacitor, and the avoided crossing of the circuit with the edge mode gives the edge weight in units of $e^2/h$, with $h$ the Planck constant. This reading is a response and not a fluctuation, so it does not require a temperature below the edge frequency (Sec.~\ref{sec:discussion}).

The floor leaves open what fixes the size of the edge weight above it, and we find that the edge itself does. For one chiral channel the edge absorption rates in the Supplemental Material of Ref.~\cite{unal2024edge} give it as a shape number, a dimensionless functional of the boundary curve and of the edge velocity along it. We write it as the Dirichlet energy of the boundary parametrized by the edge flight time, the time that an edge excitation takes to reach each point (Sec.~\ref{sec:shape}), and evaluate it in closed form for a square, $28\zeta(3)/\pi^3$ with $\zeta$ the Riemann zeta function. At a single edge velocity its terms are the absorption peaks that the edge theory gives a droplet of arbitrary shape~\cite{cano2013microwave}. Elongated lattice flakes, whose edge velocity differs from side to side, follow it without any fitted parameter. For concentric circular edges the same rates add the contributions of several channels~\cite{unal2024edge}, and in the edge theory of a droplet of arbitrary shape each edge mode contributes a series of peaks of its own~\cite{cano2013microwave}. On lattice flakes, where no edge theory is assumed, we find that every further in-gap channel accordingly adds its own shape number, traced at its own velocities, while leaving the Hall weight at the net chirality. This holds for counter-propagating pairs that run around the whole perimeter, and for the two co-propagating channels of a band with $|C|=2$ in the cases that the extrapolation resolves. A screened interaction between counter-propagating channels lowers the edge weight by an amount that agrees with a harmonic theory of the coupled channels (Sec.~\ref{sec:coupled}), and in the edge theory an unscreened Coulomb interaction lowers it by a shortfall that is consistent with an inverse logarithm of the flake size.

The inequality between the edge weight and the magnitude of the in-gap Hall weight holds transition by transition, so the calculations do not test it. They answer three questions that it leaves open, whether the in-gap Hall weight tends to $-C$, where the edge weight sits in frequency, and how large it is. Because no accessible flake separates the edge and bulk scales by decades, we bound the edge weight by sums over the edge and bulk sectors of the spectrum and by the drop of the circuit reading between the two scales, both extrapolated to large flakes, check the static variance by a projector route, and fix the size of the edge weight by an analytic edge theory.

Section~\ref{sec:theory} derives the floor and the circuit reading. Section~\ref{sec:model} gives the lattice model and the numerical routes, and Sec.~\ref{sec:results} the edge weight of the lattice flakes and its place in frequency. Section~\ref{sec:excess} treats what sets the edge weight above the floor, namely the shape of the boundary, the number of edge channels and their coupling. Section~\ref{sec:discussion} relates the results to earlier work and estimates the readout for measured disks. A reader interested in the experiment can go from Sec.~\ref{sec:theory} to Sec.~\ref{sec:discussion} directly.

\begin{figure}[tp]
\includegraphics[width=\textwidth]{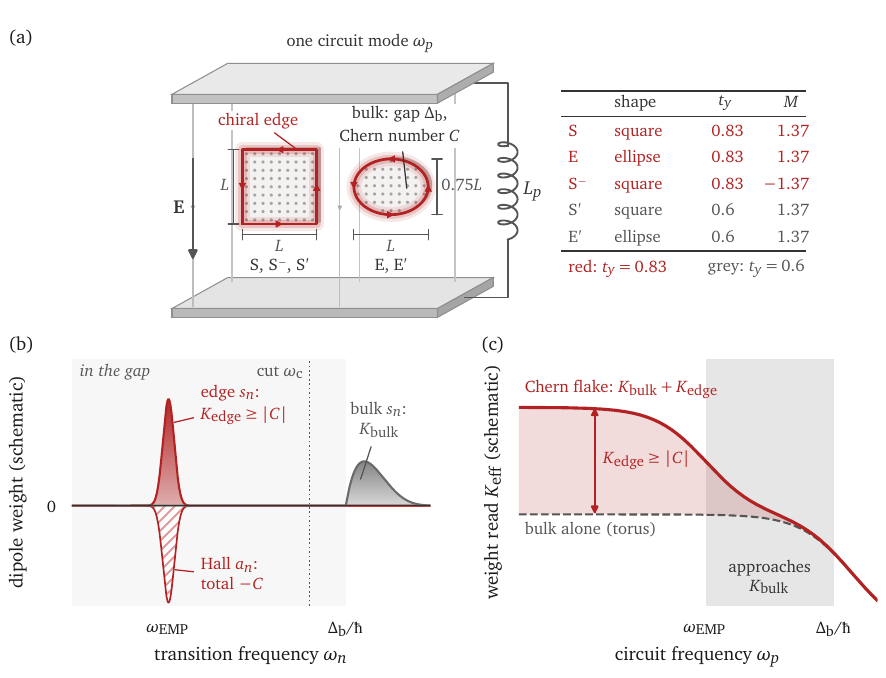}
\caption{\label{fig:schematic}Where the dipole weight of a Chern flake sits, and what a circuit reads (schematic). (a)~A circuit mode of frequency $\omega_p$ couples through the uniform field $\mathbf E$ of its capacitor to the total dipole of a Chern flake with bulk gap $\Delta_\mathrm{b}$ and Chern number $C$, a square of $L\times L$ sites or an ellipse cut from an $L\times0.75L$ rectangle. The red band is the chiral edge, and the table lists the configurations used throughout, in red for $t_y=0.83$ and grey for $t_y=0.6$, the colour code of the later figures. (b)~Dipole weight against transition frequency $\omega_n$. The in-gap edge sector near the edge-magnetoplasmon frequency $\omega_\mathrm{EMP}$ carries the symmetric weight $K_\mathrm{edge}=\sum s_n\geq|C|$ (red) and the Hall weight $\sum a_n=-C$ (hatched), and the bulk sector above $\Delta_\mathrm{b}/\hbar$ carries the bulk quantum weight $K_\mathrm{bulk}$ (grey). The dotted line is the sector cut $\omega_\mathrm{c}$. (c)~Circuit reading $K_\mathrm{eff}=\sum_n s_n\omega_n/(\omega_n+\omega_p)$ for the Chern flake (red) and the bulk alone (dashed). A circuit slower than $\omega_\mathrm{EMP}$ reads $K_\mathrm{bulk}+K_\mathrm{edge}$, and in the grey window between $\omega_\mathrm{EMP}$ and $\Delta_\mathrm{b}/\hbar$ the reading approaches $K_\mathrm{bulk}$.}
\end{figure}

\section{Edge and bulk sectors of the polarization fluctuation}
\label{sec:theory}

\subsection{Dipole weights and the dichroic sum rule}

Take a flake of area $A$ and bulk gap $\Delta_\mathrm{b}$ coupled to a uniform field through its total dipole $\hat{\mathbf P}=e\sum_i\mathbf r_i$, where $e$ is the electron charge and $\mathbf r_i$ are the electron positions [Fig.~\ref{fig:schematic}(a)].
For each excited state $|n\rangle$ at energy $\hbar\omega_n$ above the ground state $|0\rangle$, the dipole matrix element $\mathbf P_n=\langle n|\hat{\mathbf P}|0\rangle$ has $x$ component $P_{x,n}$, $y$ component $P_{y,n}$, right-circular component $P_{+,n}=(P_{x,n}+\mathrm iP_{y,n})/\sqrt2$ and left-circular component $P_{-,n}=(P_{x,n}-\mathrm iP_{y,n})/\sqrt2$.
For each transition we define the symmetric weight $s_n$ and the antisymmetric weight $a_n$ as
\begin{equation}
\begin{aligned}
s_n&=\frac{2\pi}{e^2A}\bigl(|P_{+,n}|^2+|P_{-,n}|^2\bigr),\\
a_n&=\frac{2\pi}{e^2A}\bigl(|P_{+,n}|^2-|P_{-,n}|^2\bigr),
\end{aligned}
\label{eq:weights}
\end{equation}
where $s_n=(4\pi/e^2A)(|P_{x,n}|^2+|P_{y,n}|^2)/2$ is therefore the average of the two linear absorption strengths and $a_n=(4\pi/e^2A)\,\mathrm{Im}(P_{x,n}P_{y,n}^{*})$ is the circular dichroism of the transition.
Summed over all excitations, $s_n$ gives the static polarization variance per area in units of $e^2/4\pi$. For a large periodic insulator, which has no edge, this sum reduces to the trace of the integrated quantum metric, the polarization fluctuation of the ground state~\cite{souza2000polarization}. For a band insulator the integrated quantum metric is the quantum weight~\cite{onishi2025weight}. For a bounded sample the sum also depends on the boundary.
Because the two Cartesian components of $\hat{\mathbf P}$ commute, the total antisymmetric weight vanishes identically:
\begin{equation}
\sum_{n\neq0}a_n=\frac{2\pi}{\mathrm{i}e^2A}\,\langle0|[\hat P_y,\hat P_x]|0\rangle=0 .
\label{eq:sumrule}
\end{equation}
This is the vanishing Hall conductivity of an isolated sample with boundaries~\cite{unal2024edge}.

A cut frequency $\omega_\mathrm{c}$ is set between the in-gap edge band and the bulk gap $\Delta_\mathrm{b}/\hbar$. The bulk sector, $\omega_n>\omega_\mathrm{c}$, carries the quantized circular dichroism of the filled band, $\sum a_n=C$ in our sign convention~\cite{tran2017heating}, and Eq.~\eqref{eq:sumrule} then forces the edge sector, $\omega_n<\omega_\mathrm{c}$, to carry $-C$~\cite{unal2024edge}. We call the $\sum a_n$ of a sector its Hall weight. In a flake of linear size $L$, both statements hold up to corrections of relative order $1/L$ that arise from transitions between edge states and bulk bands and from the finite width of the edge.

\subsection{The edge floor}

Each transition satisfies $|a_n|\leq s_n$, because $|P_{+,n}|^2$ and $|P_{-,n}|^2$ are not negative. Summing over the edge sector gives
\begin{equation}
K_\mathrm{edge}\equiv\sum_{\omega_n<\omega_\mathrm{c}}s_n\;\geq\;\Bigl|\sum_{\omega_n<\omega_\mathrm{c}}a_n\Bigr|=|C|+O(L^{-1}).
\label{eq:floor}
\end{equation}
The inequality is saturated only when every edge transition is fully circular with the same handedness, $P_{-,n}=0$ for all $n$ or $P_{+,n}=0$ for all $n$. On a disk at constant edge velocity the lowest chiral mode alone carries the weight, and $K_\mathrm{edge}=|C|$. An anisotropic boundary or a direction-dependent edge velocity generically mixes the two circular components and raises $K_\mathrm{edge}$ above $|C|$. Saturation requires only that the boundary coordinate $x+\mathrm iy$, parametrized by the edge flight time, contain no negative harmonics (Appendix~\ref{sec:cb}). It is therefore not confined to a disk. In flower-shaped quantum Hall droplets with more than two petals each edge absorption peak involves a single Fourier component of the boundary and does not depend on the angle of a linear polarization~\cite{oblak2024anisotropic}, which we read as a circular response of each peak. Peaks of opposite handedness both add to the symmetric weight, so such a boundary does not in general saturate the floor, and a square traced at constant velocity does not (Sec.~\ref{sec:shape}). For a droplet, the ratio of the two circular edge absorption rates measures its anisotropy, while their difference stays quantized~\cite{unal2024edge}.

The edge transitions lie close to the edge-magnetoplasmon frequency $\omega_\mathrm{EMP}$, the frequency at which an edge excitation circles the flake~\cite{volkov1988edge}, which falls as the inverse perimeter. Although the edge is a line, $K_\mathrm{edge}$ remains finite per unit area. A mode running along the perimeter displaces charge across the whole flake, so its dipole matrix element scales as $L$ and its contribution to the variance per area, $|P_n|^2/A$, stays finite as $L\to\infty$. The boundary term of the localization tensor of a Chern flake is extensive by the same counting, each of the order of $L$ boundary cells contributing a term of order $L$~\cite{marrazzo2019local}. The argument uses only the commutator, the bulk Chern number and positivity, and it assumes zero temperature and a chemical potential $\mu$ inside the bulk gap. Provided a cut frequency still separates the edge transitions from the bulk, we expect it to carry over to interacting systems with the many-body Chern number, defined through twisted boundary conditions~\cite{niu1985quantized}, in place of $C$, since the Hall conductivity of a correlated insulator, and the dichroic response to a circular drive of its twisted boundary conditions, are also set by that number~\cite{goldman2024relating}. On the lattice flakes an unscreened interaction lowers the in-gap Hall weight at the sizes that we reach (Sec.~\ref{sec:extra}), and whether it returns to $-C$ on larger flakes is open.

\subsection{What the circuit reads}

To see which part of the variance a circuit measures, couple the flake to one circuit mode of frequency $\omega_p$, with annihilation operator $b$. In the dipole gauge the total Hamiltonian $H$ is
\begin{equation}
H=H_\mathrm{el}+\hbar\omega_p b^\dagger b+g\,(b+b^\dagger)\,\hat P_{\hat e}+\frac{g^2}{\hbar\omega_p}\,\hat P_{\hat e}^{\,2},
\label{eq:hamiltonian}
\end{equation}
where $H_\mathrm{el}$ denotes the electronic Hamiltonian, $\hat P_{\hat e}$ is the fluctuation of the dipole component along the field direction $\hat e$, $g$ is a coupling strength in units of energy per dipole, and the last term is the dipole self-energy. At second order in $g$, the ground-state energy shift $\Delta E$ from this coupling is
\begin{equation}
\Delta E=\frac{g^2}{\hbar}\sum_{n\neq0}|P_{\hat e,n}|^2\Bigl(\frac{1}{\omega_p}-\frac{1}{\omega_n+\omega_p}\Bigr),
\label{eq:shift}
\end{equation}
where $P_{\hat e,n}$ denotes the matrix element of $\hat P_{\hat e}$. The average $\overline{\Delta E}$ of that shift over two orthogonal field directions defines the weight read by the circuit, the effective quantum weight $K_\mathrm{eff}$,
\begin{equation}
K_\mathrm{eff}(\omega_p)\equiv\frac{4\pi}{e^2A}\,\frac{\hbar\omega_p\,\overline{\Delta E}}{g^2}=\sum_{n\neq0}s_n\,\frac{\omega_n}{\omega_n+\omega_p}.
\label{eq:keff}
\end{equation}

For $\omega_p\to0$ every transition counts and $K_\mathrm{eff}$ reduces to the static polarization variance, $K_\mathrm{bulk}+K_\mathrm{edge}$ [Fig.~\ref{fig:schematic}(c)]. In the window $\omega_\mathrm{EMP}\ll\omega_p\ll\Delta_\mathrm{b}/\hbar$ each edge term is suppressed by $\omega_n/\omega_p$ while each bulk term is left nearly unchanged, so $K_\mathrm{eff}\to K_\mathrm{bulk}$, and the drop of the reading across that window, \begin{equation}
\Delta K\equiv K_\mathrm{eff}(0)-K_\mathrm{eff}(\omega_\mathrm{EMP}\ll\omega_p\ll\Delta_\mathrm{b}/\hbar),
\label{eq:dk}
\end{equation} is therefore $K_\mathrm{edge}\geq|C|$ for large flakes.

The edge weight $K_\mathrm{edge}$ is the weak-coupling coefficient of the reading, which a circuit measures when the circuit capacitance is large relative to the capacitance that the edge adds. Because the edge modes of the chiral-boson edge theory (Sec.~\ref{sec:model}) are harmonic, the coupled problem can be solved exactly for them (Appendix~\ref{sec:coupling}). For one transition the exact shift is
\begin{equation*}
\Delta E_n=\tfrac12\hbar(\omega_n+\omega_p)\bigl[(1+\eta_n)^{1/2}-1\bigr],\qquad \eta_n=\frac{4g^2|P_{\hat e,n}|^2\omega_n}{\hbar^2\omega_p(\omega_n+\omega_p)^2},
\end{equation*}
where $\eta_n$ denotes the coupling parameter. Hence each term of Eq.~\eqref{eq:keff} is multiplied by $2[(1+\eta_n)^{1/2}-1]/\eta_n\leq1$, and Eq.~\eqref{eq:shift} holds for $\eta_n\ll1$. In the static regime, $\omega_n\gg\omega_p$, $\eta_n$ is the ratio of the capacitance that the polarizable edge adds to the circuit to the circuit capacitance itself. For the edge sector $|P_{\hat e,n}|$ grows as $L$ while $\omega_n$ falls as $1/L$. In that regime a fixed coupling, that of a fixed capacitor, therefore makes $\eta_n$ grow as $L^3$, and once it is large the edge share of the shift per unit area falls as $L^{-3/2}$. A capacitor whose gap is scaled with the flake has a coupling that falls as $1/L$ at fixed circuit capacitance (Appendix~\ref{sec:coupling}), and $\eta_n$ then grows only as $L$.

These statements hold for a readout averaged over two orthogonal circuit modes. A single linearly polarized mode along a field direction $a$ reads one diagonal component, the axis edge weight $K_a=(4\pi/e^2A)\sum_{\omega_n<\omega_\mathrm{c}}|P_{a,n}|^2$, where $P_{a,n}$ is the component of $\mathbf P_n$ along $a$, and the average of the axis weights over $a=x$ and $a=y$ is $K_\mathrm{edge}$. For one axis, positivity by itself gives no floor. Since $|a_n|\leq(4\pi/e^2A)|P_{x,n}||P_{y,n}|$, the Cauchy--Schwarz inequality bounds only the geometric mean of the two axis weights $K_x$ and $K_y$, $(K_xK_y)^{1/2}\geq|C|+O(L^{-1})$. This is the edge counterpart of the bound of the quantum-metric determinant by the Berry curvature~\cite{roy2014band}. At least one axis therefore reads at least $|C|$.

The edge term lies inside the adiabatic regime that Onishi and Fu assume.
\begin{locked}This does not move the adiabatic condition of Ref.~\cite{onishi2026vacuum}, since for a flake its $\Delta$ is already of order $\hbar\omega_\mathrm{EMP}$ and the authors state that their theory ``breaks down when the gap is smaller than the circuit frequency''~\cite{onishi2026vacuum}.\end{locked}
Within that condition the edge term changes what the force measures, which becomes $K_\mathrm{bulk}+K_\mathrm{edge}$ rather than $K_\mathrm{bulk}$.

\section{Model and numerical method}
\label{sec:model}

\subsection{Lattice model and flakes}

Eqs.~\eqref{eq:floor} and \eqref{eq:keff} are tested on open flakes of the two-orbital Chern lattice model of Qi, Wu and Zhang~\cite{qi2006qwz}, the Bloch Hamiltonian of which at wave vector $\mathbf k$ is $\mathbf d(\mathbf k)\cdot\bm\sigma$, with $\bm\sigma$ the Pauli matrices acting on the orbitals and
\begin{equation}
\mathbf d=\bigl(t_x\sin k_x+b,\;t_y\sin k_y,\;M-t_x\cos k_x-t_y\cos k_y\bigr).
\label{eq:qwz}
\end{equation}
Here $t_x$ and $t_y$ are the hoppings along the two lattice directions, $b$ is an offset and $M$ is the mass.
The hoppings are set to $t_x=1$ and $t_y=0.83$ or $0.6$, and the offset $b=0.11$ removes the remaining mirror symmetries. The mass $M=\pm1.37$ gives $|C|=1$ while $M=2.61$ gives a trivial insulator.
The anisotropy makes the edge velocity depend on direction, so that $K_\mathrm{edge}$ can exceed $|C|$.
Flakes are squares of $L\times L$ sites and ellipses cut from an $L\times0.75L$ rectangle, with $L$ up to 80 for the elliptical flake at $t_y=0.83$ in the main series (the sizes of Table~\ref{tab:conv}), which then has 3760 sites.
Five configurations are sampled, labeled S (square, $t_y=0.83$, $M=1.37$), E (ellipse, $t_y=0.83$, $M=1.37$), S$^{-}$ (square, $t_y=0.83$, $M=-1.37$), S$^\prime$ (square, $t_y=0.6$) and E$^\prime$ (ellipse, $t_y=0.6$), each of them taken to five chemical potentials between $\pm0.3877\Delta_\mathrm{b}$. The configuration S$^{-}$ reproduces the magnitudes of S exactly, with every Hall weight reversed in sign. The two outermost chemical potentials are mapped onto each other by the symmetry $\mu\to-\mu$ and give identical results, so the sample contains four independent configurations at four independent chemical potentials.
The torus bulk weight is $K_\mathrm{bulk}=1.236$ at $t_y=0.83$ and 1.296 at $t_y=0.6$.

\subsection{Spectral decomposition}

Dense diagonalization of each flake gives every $s_n$ and $a_n$ of the particle--hole transitions, and Eq.~\eqref{eq:keff} is thus evaluated exactly, with no sector cut, on a fine logarithmic grid of circuit frequencies and interpolated between grid points, as detailed in Appendix~\ref{app:model}. The sector sums of Eq.~\eqref{eq:floor} use a cut at $0.95\Delta_\mathrm{b}$. When the cut moves from $0.7\Delta_\mathrm{b}$ to $\Delta_\mathrm{b}$, the in-gap symmetric weight changes by at most 0.015 for the largest flakes [Fig.~\ref{fig:scans}(a)]. We take the edge scale $\omega_\mathrm{EMP}$ of each flake as the in-gap transition carrying the largest symmetric weight. In the free-fermion model this is a single-particle edge excitation, the step between neighbouring levels of the chiral edge band, with a frequency falling as the inverse linear size. It has neither a magnetic field nor a Coulomb restoring force. We keep the symbol of the edge magnetoplasmon for it because an interaction turns this transition into that collective mode (Sec.~\ref{sec:extra}), and because the edge magnetoplasmon sets the edge scale of a device.

\subsection{Projector route and analytic edge theory}

A second route, the projector route, avoids the eigendecomposition. The ground-state projector, built from a Newton iteration for the matrix sign function, gives the static variance and the local Chern marker, which Bianco and Resta define directly from the projector~\cite{bianco2011marker}.
The interior marker of the largest flakes is at least 0.9992 in magnitude.
The third route is analytic and needs no diagonalization of the flake, since a chiral-boson edge theory~\cite{wen1990chiral}, built with the direction-dependent edge velocity of semi-infinite ribbons, gives the size of $K_\mathrm{edge}$, bounded below by $|C|$ by construction.
Consider a $\nu=1$ chiral edge on a closed boundary, parametrized by the flight time $\tau$ with $d\tau=ds/v$, where $s$ is the arc length and $v$ the edge velocity at $\mu$. Its density harmonics $n_m$, labelled by an integer $m$, obey $[n_m,n_{-m}]=\nu m$, with $n_{-m}$ the harmonic of opposite index, and commute with all other harmonics. The dipole components are $P_x=e\sum_m x_m n_m$ and $P_y=e\sum_m y_m n_m$, with $x_m$ and $y_m$ the Fourier coefficients defined below, so the variance of $P_x$ is $e^2\nu\sum_{m>0}m|x_m|^2$, likewise for $P_y$, and the edge dipole weight is
\begin{equation}
K_\mathrm{edge}=\frac{2\pi\nu}{A}\sum_{m>0}m\left(|x_m|^2+|y_m|^2\right)\geq\nu,
\label{eq:kb}
\end{equation}
where $x_m=T^{-1}\int_0^{T}x(\tau)\,e^{-2\pi\mathrm{i}m\tau/T}d\tau$ and $y_m$ likewise are the Fourier coefficients of the boundary coordinates over one period $T$ of $\tau$, and $m$ labels the edge harmonics.
Equality holds when $x+\mathrm{i}y$ contains no negative harmonics in $\tau$, as for a disk at constant velocity. Non-circular flakes and direction-dependent velocities generically violate that condition and give $K_\mathrm{edge}>\nu$. At a single edge velocity the terms of Eq.~\eqref{eq:kb} are the weights of the microwave absorption peaks of a quantum Hall droplet of arbitrary shape, which Ref.~\cite{cano2013microwave} obtained from the same edge theory for one field direction. For a droplet whose edge follows an equipotential, Eq.~\eqref{eq:kb} equals $\nu(S_{+}+S_{-})/(S_{+}-S_{-})$, with $S_{+}$ the edge absorption rate of one circular polarization and $S_{-}$ the rate of the other, both from Ref.~\cite{unal2024edge} [Eq.~(S54) of its Supplemental Material] with the action-angle coordinate of the confinement in place of $\tau$, and $S_{+}-S_{-}=|C|A/\pi$.
Appendix~\ref{sec:cb} gives the numerical parameters of this route.

\subsection{Window estimator}

The circuit reading develops no flat plateau at any accessible size, because the edge leakage into $K_\mathrm{eff}$, the edge weight still read, falls as $\omega_\mathrm{EMP}/\omega_p$ while the bulk loss, the bulk weight lost from the reading, grows as $\hbar\omega_p/\Delta_\mathrm{b}$.
We therefore read the drop of Eq.~\eqref{eq:dk} in a scaling window $\omega_p^{*}=\kappa\,(\omega_\mathrm{EMP}\Delta_\mathrm{b}/\hbar)^{1/2}$, with $\kappa$ a dimensionless window parameter, in which both ratios vanish as $L^{-1/2}$.
For each flake, the window drop $D(L,\kappa)=K_\mathrm{eff}(0)-K_\mathrm{eff}(\omega_p^{*})$ is read from the exact curve and extrapolated to $L\to\infty$ as a quadratic in $L^{-1/2}$.
Because Eq.~\eqref{eq:floor} forces a converged window drop to be at least $|C|$, this estimate checks the estimator and does not test the floor. Its variants and its acceptance criterion are in Appendix~\ref{app:window}.

\section{The edge weight at the floor and across the edge scale}
\label{sec:results}

\subsection{Edge-sector weight at and above the floor}
\label{sec:sectorsums}

\begin{figure}[tbp]
\includegraphics[width=\textwidth]{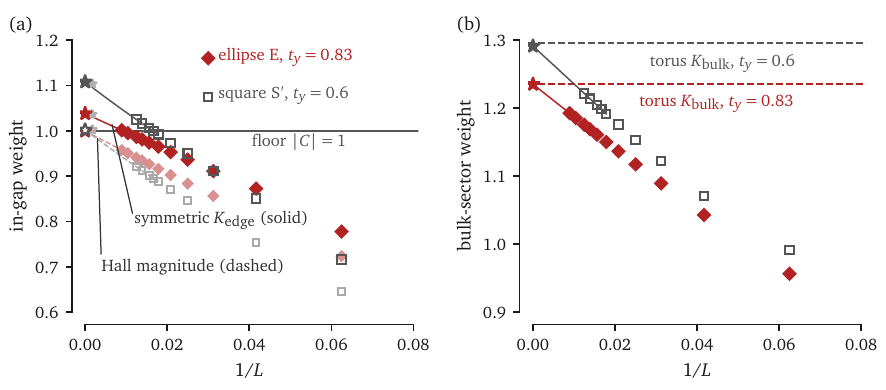}
\caption{\label{fig:sectors}The edge sector carries a symmetric weight at or above the floor. (a)~In-gap symmetric weight $K_\mathrm{edge}$ (large markers) and Hall magnitude $|\sum a_n|$ (small markers), with the sector cut at $0.95\Delta_\mathrm{b}$, against $1/L$ for the ellipse E at $t_y=0.83$ (red) and the square S$^{\prime}$ at $t_y=0.6$ (dark grey), $M=1.37$, $\mu=0.0123\Delta_\mathrm{b}$. Lines are quadratic fits over the five largest flakes (up to $L=112$ and 80), with their extrapolations as coloured stars. Grey stars repeat the fit over a smaller size set (up to $L=64$ and 56). The thin line is the floor $|C|=1$. (b)~Bulk-sector symmetric weight of the same flakes, fitted the same way, and the torus bulk weight $K_\mathrm{bulk}$ of each anisotropy (dashed).}
\end{figure}

The sector sums show, without any circuit, that the in-gap Hall weight tends to $-C$ [Fig.~\ref{fig:sectors}(a)]. Because the in-gap symmetric weight is at least the in-gap Hall magnitude at any size and for any cut, transition by transition, this turns that identity into the floor of Eq.~\eqref{eq:floor}. At finite size the in-gap Hall weight falls short of $|C|$ by the expected $1/L$ correction. The largest flakes of the main series carry an in-gap symmetric weight of 0.96--1.03 against an in-gap Hall weight of magnitude 0.89--0.94, giving a ratio above 1.04 everywhere. Extrapolated to $L\to\infty$ as a quadratic in $1/L$ across the five largest flakes of each configuration in the main and larger-flake series (Appendix~\ref{sec:round2}), the in-gap Hall weight has magnitude 0.998--1.007 for all sampled flakes and chemical potentials. The edge sector therefore carries $-C$ as the edge--bulk cancellation requires. The extrapolated in-gap symmetric weight is 1.04--1.13, and the ratio of symmetric to Hall weight is at least 1.037 in every case.

The excess of the extrapolated symmetric weight above $|C|$ is less robust than the floor. Repeating the fit over two smaller size sets, the pilot sizes, which stop at $L=48$ to 64, and the sizes of Table~\ref{tab:conv}, shifts the symmetric weight by up to 0.12 and the Hall weight by up to 0.014. At $t_y=0.6$ that shift is not monotone in the largest size, and flakes up to $L=104$ for the square and 144 for the ellipse do not settle it. At $\mu=0.0123\Delta_\mathrm{b}$ the last three size windows give an excess of 0.101--0.104 for the square and 0.051--0.067 for the ellipse, but their changes reverse sign. Over all five chemical potentials they span 0.086--0.136 and 0.035--0.075. The symmetric weight of these largest flakes already exceeds $|C|$ at every $\mu$ (Appendix~\ref{sec:round3}). The sector sums therefore establish the floor and the Hall weight, leaving the size of the excess to the analytic route of Sec.~\ref{sec:routes}.

The bulk sector returns the torus [Fig.~\ref{fig:sectors}(b)]. Its symmetric weight, extrapolated to $L\to\infty$ over the same five largest flakes, agrees with the torus bulk weight to 0.0015 for every flake at $t_y=0.83$ and to 0.0085 for all sampled flakes, with the larger residuals belonging to the more anisotropic flakes. Over the pilot sizes alone the residual reaches 0.038. The static polarization variance of a Chern flake is therefore the bulk quantum weight plus an in-gap term of at least $|C|$, as Eq.~\eqref{eq:floor} predicts for the edge and bulk sectors taken together.

\subsection{Where the edge weight sits in frequency}

\begin{figure}[tbp]
\includegraphics[width=\textwidth]{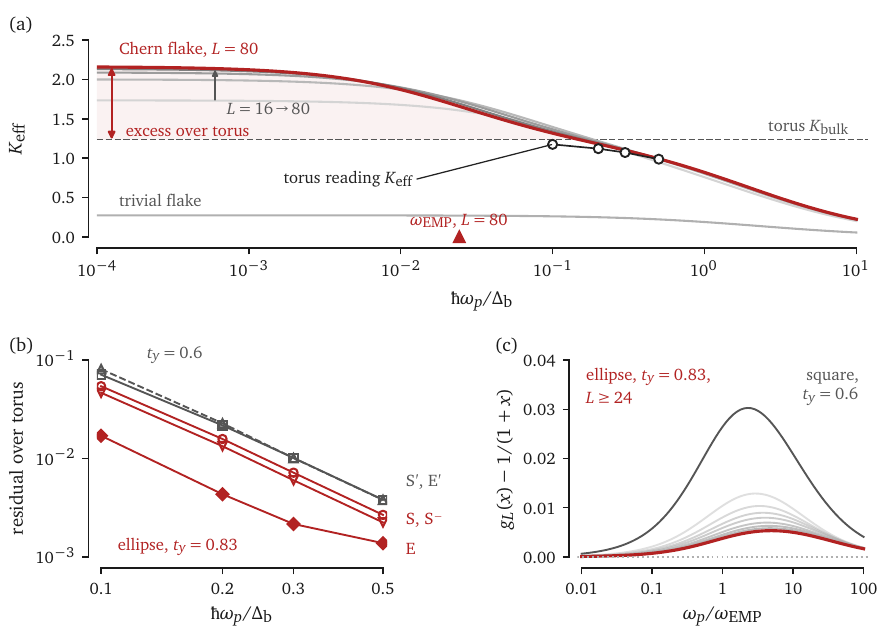}
\caption{\label{fig:keff}The edge weight leaves the circuit reading above the edge scale $\omega_\mathrm{EMP}$, the frequency of the in-gap transition of largest symmetric weight. (a)~Circuit reading $K_\mathrm{eff}$ of Eq.~\eqref{eq:keff} against $\hbar\omega_p/\Delta_\mathrm{b}$ for elliptical Chern flakes ($t_y=0.83$, $M=1.37$, $\mu=0.0123\Delta_\mathrm{b}$) with $L=16$ to 64 (grey, darkening with $L$) and $L=80$ (red), and for a trivial square flake with $L=40$. The dashed line is the torus bulk weight $K_\mathrm{bulk}$ and the open circles are the torus reading, both computed with periodic boundaries. The arrow marks the excess of the $L=80$ reading over $K_\mathrm{bulk}$, and the triangle marks $\hbar\omega_\mathrm{EMP}/\Delta_\mathrm{b}$ of that flake. (b)~Residual of the size-extrapolated flake reading over the torus reading for the five Chern configurations of Fig.~\ref{fig:schematic}(a). (c)~Deviation of the normalized in-gap reading $g_L(x)=R_L(x\omega_\mathrm{EMP})/R_L(0)$, with the in-gap reading $R_L(\omega_p)=\sum_n s_n\omega_n/(\omega_n+\omega_p)$ over the transitions below $0.95\Delta_\mathrm{b}/\hbar$, from the single-mode form $1/(1+x)$, for the ellipse at $t_y=0.83$ with $L=24$ to 80 (red for $L=80$) and the square at $t_y=0.6$ with $L=64$ (dark grey).}
\end{figure}

A slow circuit reads nearly twice the bulk weight. At $\omega_p\to0$ the extrapolated Chern-flake reading is $K_\mathrm{eff}=2.27$--2.32 for $t_y=0.83$, against the torus bulk weight $K_\mathrm{bulk}=1.24$, and it falls onto the bulk curve as $\omega_p$ rises [Fig.~\ref{fig:keff}(a)]. The step sits at the edge scale, which falls as the inverse perimeter [Fig.~\ref{fig:window}(a)], and at the largest flakes $\hbar\omega_\mathrm{EMP}$ is 0.024--0.052 of the bulk gap, with the lower end belonging to the less anisotropic flakes.

The edge weight leaves the reading once the circuit outpaces the edge [Fig.~\ref{fig:keff}(b)].
Because $\omega_\mathrm{EMP}\to0$ as $L\to\infty$, the size-extrapolated flake reading at any fixed $\omega_p$ should equal the torus bulk curve, and the residual excess measures how well the extrapolation in $1/L$ converges at that frequency.
At $\hbar\omega_p=0.3\Delta_\mathrm{b}$ the residual is 0.002--0.010 over all configurations, and at $0.5\Delta_\mathrm{b}$ it is 0.001--0.004.
At $0.1\Delta_\mathrm{b}$ the residual rises to 0.017--0.081, being largest for the more anisotropic flakes, whose edge scale at the largest size lies closest to the circuit frequency.
The trivial flake follows the torus curve to within $4\times10^{-5}$ at every one of these frequencies.
Within these residuals the high-frequency reading of a Chern flake is that of the bulk, and the weight of at least $|C|$ that the static reading carries is gone.

Measured in units of the edge scale, the in-gap part of the reading falls off as $1/(1+x)$ with $x=\omega_p/\omega_\mathrm{EMP}$ [Fig.~\ref{fig:keff}(c)].
Over every size $L\geq24$ of each Chern configuration, its value normalized to the static one varies by at most 0.014 at $x=0.3$, 1, 3 and 10, whereas read at a fixed $\hbar\omega_p/\Delta_\mathrm{b}$ it varies at least 18 times as much.
This collapse is largely forced, since one transition, the edge magnetoplasmon, carries 89--99\,\% of the in-gap weight and defines $x$. What it adds is that the remaining weight does not sit near the bulk gap, at most 2.3\,\% of the in-gap weight lying above $0.5\Delta_\mathrm{b}$.
The edge weight therefore leaves the reading by an algebraic roll-off rather than a step, so that at $\mu=0.0123\Delta_\mathrm{b}$, 26--28\,\% of it is still read at $\omega_p=3\omega_\mathrm{EMP}$ and 10--11\,\% at $10\omega_\mathrm{EMP}$.

\subsection{Window drop as a check of the estimator}

Since the edge Hall sum rule and Eq.~\eqref{eq:floor} force a converged calculation to give $\Delta K/|C|\geq1$, the window drop validates the estimator. It does not resolve the excess of $K_\mathrm{edge}$ above $|C|$, which for the elliptical flake with $t_y=0.83$ falls inside one spread of zero.
Over the sampled flakes, the minimum of the extrapolated window drop is $\Delta K/|C|=1.05\pm0.06$, attained at the elliptical flake with $t_y=0.83$ and $\mu=0.0123\Delta_\mathrm{b}$ [Fig.~\ref{fig:box}].
The spread combines the half-range of nine extrapolation variants, three values of $\kappa$ times three fit forms (Appendix~\ref{app:window}), which is 0.057 at that flake, with the largest change on adding the biggest flake anywhere, 0.015.
It is a systematic spread within one extrapolation form in $L^{-1/2}$, not a statistical error, and it does not cover extrapolation variables other than $L^{-1/2}$.
The raw $\kappa=1$ drops of the largest flakes are 0.87--0.96, all below $|C|$, so the value above the floor comes from the extrapolation [Fig.~\ref{fig:window}(b)].
The acceptance criterion, the variants and the larger flakes are in Appendix~\ref{app:window}.

\subsection{Size of the edge weight from three routes}
\label{sec:routes}

\begin{figure}[tbp]
\includegraphics[width=\textwidth]{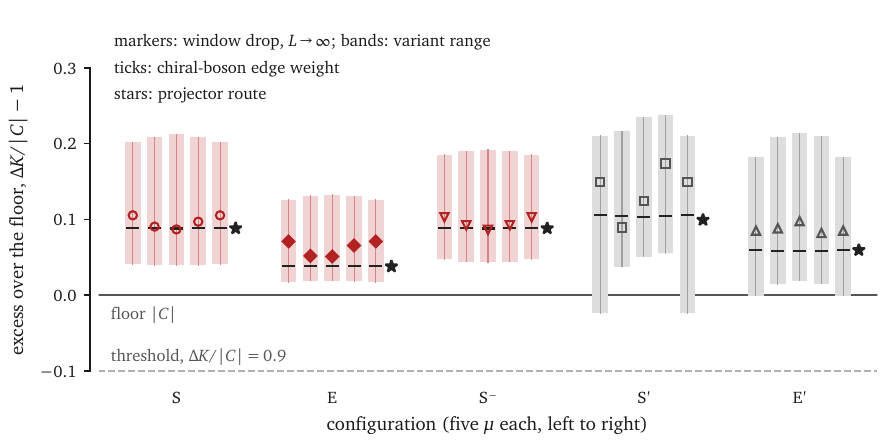}
\caption{\label{fig:box}Three routes to the edge weight. Excess over the floor, $\Delta K/|C|-1$, of the extrapolated window drop (markers, with the range of the nine extrapolation variants as bands) for each configuration at the five chemical potentials, from $\mu=-0.3877\Delta_\mathrm{b}$ to $0.3877\Delta_\mathrm{b}$ left to right. Black ticks give the analytic chiral-boson edge weight, and black stars the projector route at $\mu=0.0123\Delta_\mathrm{b}$, both minus 1. The solid line is the floor $|C|$, and the dashed line is the threshold $\Delta K/|C|=0.9$.}
\end{figure}

The three routes place every Chern configuration at or above $|C|$ [Fig.~\ref{fig:box}].
The projector route takes the static variance, extrapolated over the pilot sizes, minus the torus bulk weight at $\mu=0.0123\Delta_\mathrm{b}$. It gives 1.038--1.099 over the five configurations and 1.038 for the elliptical flake at $t_y=0.83$, and it shares the flakes with the window estimator but not its spectral decomposition or its frequency window. It evaluates the same static variance as the sector sums by other numerical means and subtracts the same torus weight, so it checks the numerics and is not an independent determination of the edge weight. The independent determination is the analytic route.
The analytic chiral-boson edge weight shares only the lattice Hamiltonian, through the ribbon edge velocities, and gives 1.0381 for that ellipse and 1.038--1.106 over all sampled flakes. The window estimate minus this route ranges from $-0.015$ to $+0.069$.

The excess above $|C|$ comes from the shape of the flake and from the direction dependence of its edge velocity. The analytic route places the edge weight of every Chern configuration above $|C|$, least for the less anisotropic ellipse and most for the more anisotropic square. At constant velocity Eq.~\eqref{eq:kb} gives 1.086 for a square, which the square flakes are, and 1.024 for an ellipse of aspect ratio 0.75. With the ribbon velocities at $\mu=0.0123\Delta_\mathrm{b}$ it gives 1.088 and 1.104 for the squares at $t_y=0.83$ and 0.6 and 1.038 and 1.058 for the ellipses. The direction-dependent velocity therefore adds 0.002 to 0.035 to the shape value. The size of the excess is resolved in this way for the flakes at $t_y=0.83$ and for the shape-law series of Sec.~\ref{sec:shape}. For the more anisotropic flakes at $t_y=0.6$ the sector-sum excess is not converged (Sec.~\ref{sec:sectorsums}), and we make no claim for its size there.
Lattice disks with isotropic hopping, $t_y=t_x$, are consistent with saturation at the 2\,\% level. Their in-gap symmetric weight extrapolates to 1.006--1.012, with an extrapolation spread of 0.011--0.023, against a chiral-boson value of 1.001--1.002. The ellipse at $t_y=0.83$, by contrast, carries an excess of 0.037--0.041 on the same estimator, with a spread of 0.002--0.005 (Appendix~\ref{sec:round3}). These spreads measure the stability of the fit between size windows and not its bias. On the disks, where the edge theory fixes the answer, the same estimator lies above it by half a per cent to one per cent, so the excess of the ellipse is resolved and its third decimal is not.

\subsection{Single axes, trivial edges and screening}
\label{sec:extra}

Resolved by axis, the edge weight follows the shape and the velocity anisotropy of the flake [Fig.~\ref{fig:round2}(a), Appendix~\ref{sec:round2}]. Extrapolated in the same way over the five largest flakes of the main series, the axis weight $K_x$ is 1.15--1.41 and $K_y$ is 0.70--1.03 over the sampled flakes and chemical potentials. For the elliptical flakes, cut with an aspect ratio of 0.75, $K_x=1.291$ and $K_y=0.780$ at $t_y=0.83$ and $K_x=1.395$ and $K_y=0.715$ at $t_y=0.6$ ($\mu=0.0123\Delta_\mathrm{b}$). Including the larger flakes moves these four values by up to 0.020. The ratio of their geometric mean to the in-gap Hall magnitude, which extrapolates to 1.001 and 0.999, is 1.002 and 1.000, so an ellipse saturates the bound on $(K_xK_y)^{1/2}$ rather than the bound on $K_\mathrm{edge}$. Anisotropic quantum Hall droplets show a perfect elliptic dichroism of their bulk response, a regime in which they respond to only one elliptical polarization~\cite{elliptic2026dichroism}, and the saturated geometric mean is its counterpart in the edge sector. At every computed size $(K_xK_y)^{1/2}$ is at least the in-gap Hall magnitude, as Cauchy--Schwarz requires. The window drop read by a single circuit mode follows the same pattern, 1.34 along $x$ and 0.76 along $y$ for the ellipse at $t_y=0.83$, with variant half-ranges of 0.09 and 0.05. A single linearly polarized mode along the short axis of an elongated flake therefore reads an edge step below $|C|$.

A trivial flake with metallic edge states also sheds low-frequency weight, here about twice that of a Chern flake, but it carries no Hall weight [Fig.~\ref{fig:round2}(b)]. We attach to every perimeter site of the trivial flake one extra orbital, coupled to its site and to its neighbors around the perimeter, which forms a non-chiral ring band lying between $-0.34\Delta_\mathrm{b}$ and $0.30\Delta_\mathrm{b}$ (Appendix~\ref{sec:round2}). With $\mu$ inside the ring band the in-gap symmetric weight extrapolates to 2.17 and the window drop to 2.18, about twice the Chern values, while the in-gap Hall weight is below $10^{-4}$ in magnitude. With $\mu$ outside the ring band but inside the bulk gap the in-gap weight at $L=56$ does not exceed 0.0031, and the step is gone. The same ring attached to the Chern flake leaves its in-gap Hall weight at magnitude 0.996--1.003 for every $\mu$.

Screening by a gate-limited Coulomb interaction raises the edge frequency and, at short screening length, leaves the edge weight unchanged [Fig.~\ref{fig:device}(a)]. We treat the square flake at $t_y=0.83$ in self-consistent Hartree theory and its dipole response in time-dependent Hartree theory, the random-phase approximation, with an interaction of strength $U$ and gate screening length $\lambda$ between site densities (Appendix~\ref{sec:round2}). Exchange is left out, so the calculation tests the direct channel, the depolarizing field of the edge charge, and not the exchange renormalization of the edge velocity. For flakes of $L=12$ to 28 at $\lambda=3$ the collective edge frequency exceeds the free one by a factor of 1.12, 1.23 and 1.46 for $U=0.25$, 0.5 and 1 (in units of $t_x$), independent of size to 0.004 for $L\geq16$. At $U=1$ the factor is 1.29 and 1.65 for $\lambda=1.5$ and 6, rising with the screening length as a gate-limited Coulomb velocity does. The edge mode carrying the largest weight is circular to within $4\times10^{-4}$, the in-gap Hall weight extrapolates to magnitude 1.000--1.004 and the in-gap symmetric weight to 1.08 at every $U$, the same as for free fermions. The response with every particle--hole pair included (Appendix~\ref{sec:round2}) lowers the static reading at $L=24$ from 2.05 to 1.98 at $U=1$ and leaves the reading at $\hbar\omega_p=0.3\Delta_\mathrm{b}$ at 1.15--1.17. Its window drop at the screened edge scale extrapolates, as a quadratic in $L^{-1/2}$ over $L=12$ to 24, to 1.10 without and 1.14 with interaction. These flakes are small, and the transition space of the mode-resolved calculation (Appendix~\ref{sec:round2}) is cut at $1.5\Delta_\mathrm{b}$. Moving the cut to $2.5\Delta_\mathrm{b}$ lowers the in-gap symmetric and Hall weights of the $L=16$ flake at $U=1$ by 0.016 and 0.018.

Without gate screening, $\lambda\to\infty$, the collective edge frequency at $U=1$ exceeds the free one by a factor that grows with size, from 1.64 at $L=12$ to 1.96 at $L=28$, and below $\lambda\approx L/2$ it grows logarithmically in $\lambda$ (Appendix~\ref{sec:round3}).
With every pair included, the window drop at the screened edge scale then extrapolates to 1.10, the same as without interaction, within a spread of about 0.1 between extrapolation forms.
In the mode-resolved calculation the extrapolated in-gap symmetric and Hall weights both fall, the symmetric one by 0.025 at $U=1$ and by 0.014 at $U=0.5$. Moving the cut from 1.5 to $2.5\Delta_\mathrm{b}$ lowers them by 0.025--0.026 at $\lambda\to\infty$, so that shift is not resolved, and we claim unchanged in-gap weights at $\lambda=3$ only. A calculation with every particle--hole pair, in which the in-gap sums are taken by a contour integral of the dipole response, removes that cut (Appendix~\ref{sec:round3}). On flakes of $L=12$ to 28 both in-gap weights then lie below those of free pairs in the same mean field at every size, by an amount that shrinks as the flake grows. At $\lambda=3$ and $U=1$ the symmetric weight lies 7.2\,\% below at $L=12$ and 3.6\,\% below at $L=28$, and a quadratic in $1/L$ through the five sizes returns both weights to those of the free flake, 1.084 against 1.083 for the symmetric weight and a Hall magnitude of 1.003. This confirms the extrapolated weights of the truncated calculation. Without a gate at $U=1$ the symmetric weight lies 10.2 and 7.3\,\% below at the same two sizes and the Hall weight 9.8 and 6.9\,\%, and the same fit gives a Hall magnitude of 0.971, and 0.987 at $U=0.5$. The first misses one by more than our tolerance of 0.02. A deficit that vanishes for large flakes fits the five sizes as closely as one that survives, so we treat the two extrapolated weights without a gate as unresolved. With and without a gate the ratio of the symmetric to the Hall weight stays within 0.5\,\% of that of the free pairs at every size. We claim only that the interaction leaves this ratio in place, and whether the in-gap Hall weight of an ungated flake returns to $-C$ on large flakes is open.

Anderson disorder leaves the edge weight in place up to a disorder strength equal to the bulk gap.
We add a random on-site energy of disorder strength $W$, drawn uniformly from $[-W/2,W/2]$ on both orbitals of each site, and average over 16 realizations. For $W\leq\Delta_\mathrm{b}$ the in-gap Hall weight of the square and of the ellipse at $t_y=0.83$ stays at $-C$, extrapolating to magnitude 0.999--1.000 at $\mu=0.0123\Delta_\mathrm{b}$. The in-gap symmetric weight stays above $|C|$, within 0.003 of its clean value at that $\mu$ and within 0.013 at all five. At $\mu=0.0123\Delta_\mathrm{b}$ the window drop is unchanged to within 0.014 up to $W=2\Delta_\mathrm{b}$ (Appendix~\ref{sec:round3}).

\section{What sets the edge weight above the floor}
\label{sec:excess}

The floor is only a lower bound. Above it the edge weight is set by the edge itself, by the shape of the boundary and the edge velocities for one channel, by the number of channels, and by the interaction between them.

\subsection{Shape law}
\label{sec:shape}

\begin{figure}[tbp]
\includegraphics[width=\textwidth]{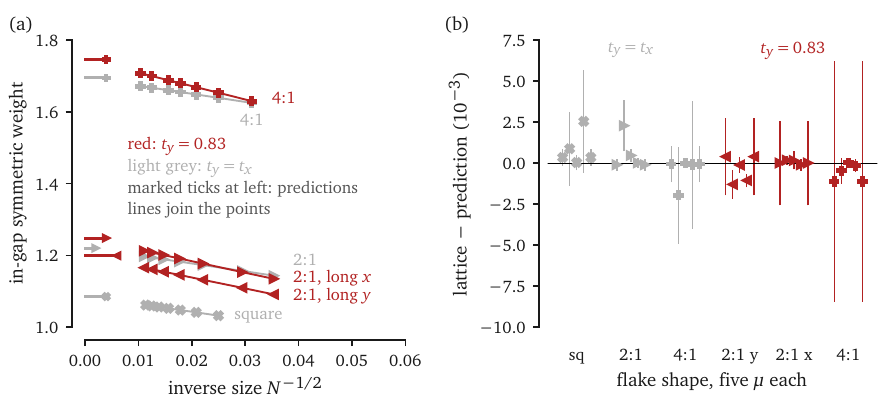}
\caption{\label{fig:shape}Shape law on lattice flakes. (a)~In-gap symmetric weight at $\mu=0.0123\Delta_\mathrm{b}$ against inverse size $N^{-1/2}$, with $N$ the number of sites, for the six series, squares and 2:1 and 4:1 rectangles at $t_y=t_x$ (light grey) and 2:1 rectangles with the long side along $x$ or $y$ and a 4:1 rectangle at $t_y=0.83$ (red). Markers give the shape, a cross for the square, right-pointing triangles for 2:1 rectangles long along $x$, left-pointing triangles for the 2:1 rectangle long along $y$ and plus signs for 4:1 rectangles. Lines join the points. Ticks at $N^{-1/2}=0$, in the colour of their series and ending in its marker, give Eq.~\eqref{eq:kb} with the ribbon velocities. (b)~Extrapolated lattice value minus that prediction for every series and the five chemical potentials, with the spread of three extrapolation windows as bars. The horizontal line marks the prediction.}
\end{figure}

For one chiral channel the edge weight is a geometric number of the boundary, which elongated lattice flakes follow without any fitted parameter [Fig.~\ref{fig:shape}]. The edge absorption rates behind it are given for an arbitrary equipotential boundary in the Supplemental Material of Ref.~\cite{unal2024edge}, and for a boundary of arbitrary shape traced at one velocity by the absorption spectrum of Ref.~\cite{cano2013microwave}.
Write the boundary as $z(\tau)=x(\tau)+\mathrm{i}y(\tau)$, parametrized by the flight time $\tau$ of Eq.~\eqref{eq:kb}, with Fourier coefficients $z_m$ over one period, labelled by the integer $m$, and let $\nu$ be the filling factor of the channel, equal to $|C|$ on the lattice flakes. Eq.~\eqref{eq:kb} is then identically
\begin{equation}
\frac{K_\mathrm{edge}}{\nu}=\frac{\sum_m|m|\,|z_m|^2}{\sum_m m\,|z_m|^2}=\frac{D[z]}{A},
\label{eq:douglas}
\end{equation}
where $D[z]=\tfrac12\int_{|w|<1}|\nabla\tilde z|^2\,d^2w$ is the Dirichlet energy of the harmonic extension $\tilde z$ of $z$ into the unit disk, with $w$ the complex coordinate of that disk, and the denominator is $A/\pi$ for a boundary traced counter-clockwise. The floor is the area inequality $D[z]\geq A$ of Douglas~\cite{douglas1931solution}, with equality only when $\tilde z$ is a conformal map, that is, when the flight time is the harmonic measure of the boundary seen from some interior point, the probability with which a random walk started at that point first reaches each part of the boundary. As a check of the code, boundaries that are images of the unit circle under polynomial conformal maps, traced at that flight time, give 1 to $10^{-15}$, as the identity requires, while the same curves at constant velocity give 1.005--1.031. Saturation therefore fixes a velocity profile for every shape rather than a shape. At constant velocity the excess is a pure shape number, $28\zeta(3)/\pi^3=1.0855$ for a square, 1.2212 and 1.6961 for rectangles of aspect ratio 2:1 and 4:1, 1.3094 for an equilateral triangle and 1.0184 for a regular hexagon. These statements hold within the hypotheses of Eq.~\eqref{eq:kb}, one chiral channel with a harmonic edge, a local edge velocity, a sharp edge, zero temperature and $\mu$ in the bulk gap, with corner and edge-width corrections of order $1/L$ on a lattice.

On lattice flakes the shape law holds for every shape, orientation and chemical potential computed.
Eq.~\eqref{eq:kb} was evaluated with the ribbon velocities of each model, and the comparison was a shape-independent weight equal to the square value. Six series were run at $M=1.37$, squares at $t_y=t_x$ and rectangles of aspect ratio 2:1 and 4:1 at $t_y=t_x$ and $t_y=0.83$, the 2:1 rectangle at $t_y=0.83$ with its long side along $x$ and along $y$, with up to 9216 sites, at the five chemical potentials of Sec.~\ref{sec:sectorsums}. The in-gap symmetric weight is extrapolated as in Sec.~\ref{sec:sectorsums}, over three windows of five consecutive sizes. Across all thirty extrapolated values the lattice lies within 0.0026 of the prediction, with a window spread of at most 0.0073 against a tolerance of 0.03, and the in-gap Hall magnitude lies within 0.0016 of 1. For the rectangles the shape-independent value misses by 0.11 to 0.67, while for the squares it coincides with the prediction and is no test. Velocity enters through the orientation. At $t_y=0.83$ and $\mu=0.0123\Delta_\mathrm{b}$ the same 2:1 rectangle gives 1.2486 with its long side along $x$ and 1.1993 with it along $y$, against predictions of 1.2484 and 1.1994, because each side contributes with its own edge velocity. The values at $\mu=\pm0.3877\Delta_\mathrm{b}$ coincide in every series, so each series tests four independent chemical potentials.

The square value bears on a published number. Moosavi and co-workers express the edge contribution to the position variance of a quantum Hall droplet as a double integral over its boundary and quote $\mathrm{Var}[x]\approx0.5183083$ for a square droplet at $\nu=1$, a value that their Supplemental Material writes as an infinite series~\cite{moosavi2025quantum}. The canonical angle that they give for the square advances linearly with the arc length along each side, which is a constant edge velocity in our terms, and the bulk term $1/4$ of the series corresponds to an area of $\pi$. With the bulk term $1/4$ removed, the edge term 0.2683083 corresponds to $K_\mathrm{edge}=1.0732$ in our units. The edge term of their variance formula, Eq.~(27) of Ref.~\cite{moosavi2025quantum}, summed term by term (Appendix~\ref{sec:square}), gives 0.2713774, equal to $7\zeta(3)/\pi^3$ and to $K_\mathrm{edge}=1.0855$. We could not reproduce the quoted value from that formula, and the two differ by 0.0031 in the edge term (Appendix~\ref{sec:square}). Lattice squares at $t_y=t_x$, with up to 7744 sites, extrapolate to within 0.0026 of 1.0855 at all five chemical potentials. A lattice flake is not a lowest-Landau-level droplet, so this shows that lattice flakes follow Eq.~\eqref{eq:kb}, and it does not by itself decide between the two published numbers.

\subsection{Counting edge channels}
\label{sec:count}

\begin{figure}[tbp]
\includegraphics[width=\textwidth]{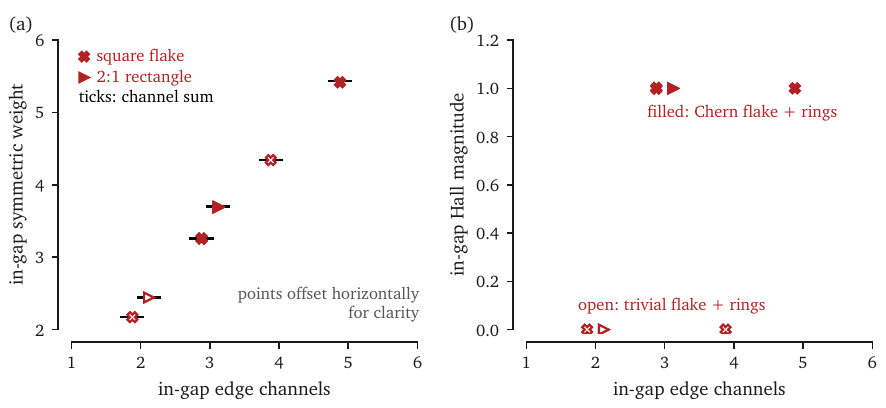}
\caption{\label{fig:count}Counting edge channels. (a)~Extrapolated in-gap symmetric weight against the number of in-gap edge channels for square flakes (crosses) and 2:1 rectangles (triangles) with ring orbitals on a trivial host (open markers) and a Chern host (filled markers), at $t_y=0.83$. Points are offset horizontally from the integer channel number for clarity. Black ticks give the sum over branches of Eq.~\eqref{eq:kb}. (b)~In-gap Hall magnitude of the same flakes.}
\end{figure}

The in-gap symmetric weight counts every edge channel, while the in-gap Hall weight counts only their net chirality [Fig.~\ref{fig:count}].
For concentric circular edges at a common radius this counting follows from the edge absorption rates of Ref.~\cite{unal2024edge} [Eq.~(S45) of its Supplemental Material], where the absorption rate $S_{+}$ of one circular polarization adds the filling factors of the right-moving modes and the absorption rate $S_{-}$ of the other adds those of the left-moving ones. For a droplet of arbitrary shape the edge theory gives each edge mode a series of absorption peaks spaced by its own velocity, so that in the clean limit the absorption spectrum counts the edge modes~\cite{cano2013microwave}. Summed over frequency, and with velocities that differ from side to side, this is the hypothesis that we test on lattice flakes.
For an edge with several in-gap branches $b$, each a Fermi crossing of an edge band, we tested the hypothesis $K_\mathrm{edge}=\sum_bD_b$, where $D_b$ is Eq.~\eqref{eq:kb} for the boundary traced with the velocity of branch $b$, together with an in-gap Hall weight equal to the number of right-moving minus left-moving branches. The branches are counted on ribbons cut along the four sides, with the ring orbitals of Sec.~\ref{sec:extra}, which add a counter-propagating pair. We call the flake that carries the rings the host, trivial or Chern. A second ring, coupled to the other host orbital with on-site energy 0.05, hopping $-0.35$ along the edge and coupling 0.25, adds a second pair. The predictions are compared with two alternatives, a weight per branch independent of shape and a second non-chiral pair that adds nothing.
The tolerance was $\max(0.03,s)$, with $s$ the extrapolation spread, and only chemical potentials at which every in-gap branch runs around the whole perimeter and the spread is below 0.05 count as tests.

Every such test agrees, from two to five in-gap channels. The 2:1 rectangles with one ring at $t_y=0.83$ give 2.4427 and 2.4437 on the trivial host against 2.4429 and 2.4440, and 3.6850 and 3.6883 on the Chern host against 3.6906 and 3.6934, where a shape-independent weight would give 2.171 and 3.259. Squares with two rings give 4.3448 and 4.3413 on the trivial host against 4.3420, and 5.4165 and 5.4173 on the Chern host against 5.4294--5.4393 and 5.4297, where a saturating second pair would give 2.171 and 3.259. The five-channel flakes lie 0.012--0.013 below the prediction, inside the tolerance. The squares with one ring agree as well, at 2.1709--2.1710 and 3.2545--3.2617. The in-gap Hall magnitude is 0.998--1.003 on every Chern host and below $3\times10^{-4}$ on every trivial one.
For a square of known edge velocities the symmetric weight in units of the square value 1.0855 is 2.000, 2.998--3.005, 3.999--4.003 and 4.990--4.991 for two to five channels. Read together with the Hall weight it gives the number of non-chiral channels of a clean edge without contacts.
The excluded cases are the Chern host with one ring at $\mu=-0.3877\Delta_\mathrm{b}$, where the counter-propagating pair exists on one side only, as a pocket holding about $L/23$ electrons, and at $\mu=0.4877\Delta_\mathrm{b}$, close to the edge of the ring band, where the spread between windows is 0.09 for squares up to $L=92$ and 0.84 for the 2:1 rectangle. Without mixing at the corners the hypothesis extends to the chiral loop plus a segment for that side, 1.3587 for the square. Squares up to $L=92$ leave the extrapolated symmetric weight unresolved, 1.29 with a spread of 0.34 between windows. A pocket of free fermions would add its weight in steps with $L$ rather than smoothly. A step is found between $L=62$ and 68, where this picture places one, but its height of 0.0452 lies outside the window of 0.015 to 0.045 that the picture allows, so it is not confirmed. Only the axis weight perpendicular to the pocket side resolves, $K_y=1.0865$ with a spread of 0.015, against 1.0981 without corner mixing and 1.3281 with complete transfer at the corners, which it excludes. Branches confined to part of the perimeter therefore lie outside the tested range of the hypothesis.

\begin{figure}[tbp]
\includegraphics[width=\textwidth]{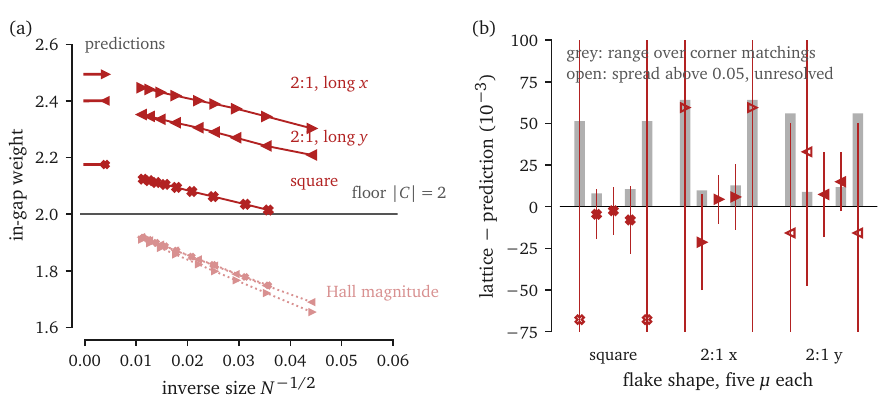}
\caption{\label{fig:c2}A Chern band with $|C|=2$ on lattice flakes. (a)~In-gap symmetric weight (dark markers) and in-gap Hall magnitude (light markers) at $\mu=0.0123\Delta_\mathrm{b}$ against inverse size $N^{-1/2}$, with $N$ the number of sites, for squares (crosses) and for 2:1 rectangles with the long side along $x$ (right-pointing triangles) and along $y$ (left-pointing triangles). Ticks at $N^{-1/2}=0$ give the sum over the two branches of Eq.~\eqref{eq:kb}, and the thin line is the floor $|C|=2$. (b)~Extrapolated lattice value minus the prediction without exchange at the corners, for each shape at the five chemical potentials, from $\mu=-0.3877\Delta_\mathrm{b}$ to $0.3877\Delta_\mathrm{b}$ left to right, with the spread of three extrapolation windows as bars. Grey bars give the range of the prediction over all corner matchings. Open markers are cases whose spread exceeds 0.05, with bars cut by the frame.}
\end{figure}

A Chern band with $|C|=2$ carries two co-propagating channels of its own, and in every case that the extrapolation resolves it follows the same counting [Fig.~\ref{fig:c2}]. We take a two-orbital model with hoppings to first and second neighbours whose lower band has a Chern number of magnitude two and a torus bulk weight of 2.153 (Appendix~\ref{app:c2}).
Its ribbons carry two edge branches of the same chirality on every side at all five chemical potentials. Two such branches can exchange at a corner, which the hypothesis leaves open, so its prediction is the range over all corner matchings. That range has no width where both branches have the same velocity on every side, as they nearly do at $\mu=0.0123\Delta_\mathrm{b}$, and the test there is that of two copies of the single-channel shape law. It is widest at $\mu=\pm0.3877\Delta_\mathrm{b}$, where the faster branch outruns the slower by a factor of 1.55 on the sides along $x$ and 2.32 on those along $y$. The predictions use the ribbons alone.
At $\mu=0.0123\Delta_\mathrm{b}$ the squares extrapolate to 2.1728 against 2.1753, where one branch alone would give 1.088. The 2:1 rectangles give 2.4986 with the long side along $x$ and 2.4078 with it along $y$, against 2.4942 and 2.4005, where the square value would give 2.175 for both.
Of the twelve independent cases, three shapes at four chemical potentials, eight agree with the prediction within the tolerance, lying within 0.021 of the range of the matchings, and four are unresolved, with an extrapolation spread above 0.05 (Appendix~\ref{app:c2}). At $\mu=\pm0.3877\Delta_\mathrm{b}$, where the two branches differ most in velocity, none of the three shapes is resolved, so the counting is not tested where the branches differ most. An attempt to resolve the remaining cases by averaging the sums over the chemical potential failed a control that we had fixed for it, and we draw no conclusion from it (Appendix~\ref{app:c2}).
The in-gap Hall magnitude extrapolates to 1.999--2.001 in every case, so the edge sector of this band carries $-C$ and its floor is two. The static reading of the squares is 4.33, again about twice the torus bulk weight.

\subsection{Coulomb edges}
\label{sec:coulomb}

\begin{figure}[tbp]
\includegraphics[width=\textwidth]{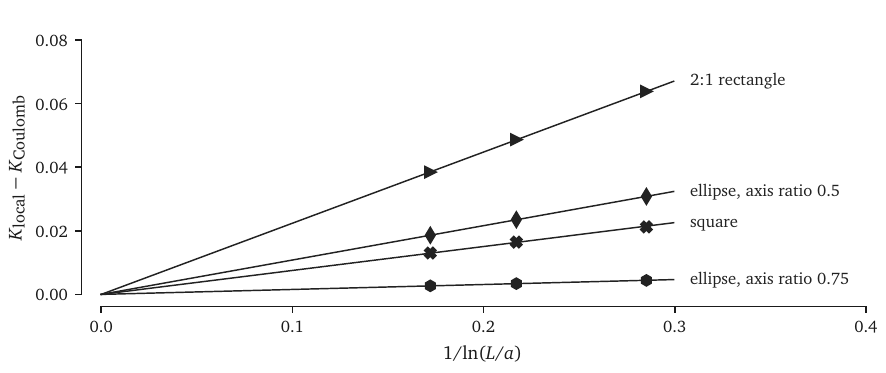}
\caption{\label{fig:coulomb}Coulomb edges in the edge theory. Shortfall of the edge weight below its value for a local edge, $K_\mathrm{local}-K_\mathrm{Coulomb}$, against $1/\ln(L/a)$ at $g_\mathrm{C}/u_0=10^3$ and $a/L=0.03$, 0.01 and 0.003, for a square (crosses), a 2:1 rectangle (triangles) and ellipses of axis ratio 0.75 (hexagons) and 0.5 (thin diamonds), each labelled at the end of its line. Lines through the origin have slope $c$, the mean of $(K_\mathrm{local}-K_\mathrm{Coulomb})\ln(L/a)$ over the three cutoffs.}
\end{figure}

An unscreened Coulomb interaction along the edge lowers the edge weight below the shape value, and in the edge theory the shortfall falls off as $1/\ln(L/a)$, with $a$ a short-distance cutoff [Fig.~\ref{fig:coulomb}].
We take one chiral branch with a local drift term $u_0$ and a kernel of strength $g_\mathrm{C}$ between a point $\bm r$ and a point $\bm r'$ of the boundary, $g_\mathrm{C}/\sqrt{|\bm r-\bm r'|^2+a^2}$, traced at constant drift velocity. The boundary is discretized into points of equal arc length, the kernel is taken to the basis of the density harmonics, and the ground-state covariance follows from the matrix formula of Eq.~\eqref{eq:coupled} below, with the harmonics in place of the branches. The results are stated with $K_\mathrm{Coulomb}$ the edge weight with this kernel and with $K_\mathrm{local}$ the edge weight without it. A disk keeps $K_\mathrm{edge}=1$ to $3\times10^{-15}$ at $g_\mathrm{C}/u_0=5$ and $a/L=0.01$, because rotation symmetry keeps each density harmonic an eigenmode, and with $g_\mathrm{C}=0$ the calculation reproduces Eq.~\eqref{eq:kb} for the disk, the square, the 2:1 rectangle and the ellipse of axis ratio 0.75. At $g_\mathrm{C}/u_0=10^3$ and $a/L=0.03$, 0.01 and 0.003 the square gives 1.0643, 1.0690 and 1.0725 against 1.0855, and the product $(K_\mathrm{local}-K_\mathrm{Coulomb})\ln(L/a)$ is 0.0744, 0.0758 and 0.0757. For the 2:1 rectangle and for ellipses of axis ratio 0.75 and 0.5 the same product is 0.223--0.224, 0.0154--0.0157 and 0.108. The weights are therefore consistent with $K_\mathrm{Coulomb}=K_\mathrm{local}-c/\ln(L/a)$ with a shape-dependent constant $c$. With three cutoffs and the constant inside the logarithm not fitted, this is a consistency statement rather than a fitted law, and it rests on the edge theory alone, without a lattice test.

\subsection{Coupled edge channels}
\label{sec:coupled}

\begin{figure}[tbp]
\includegraphics[width=\textwidth]{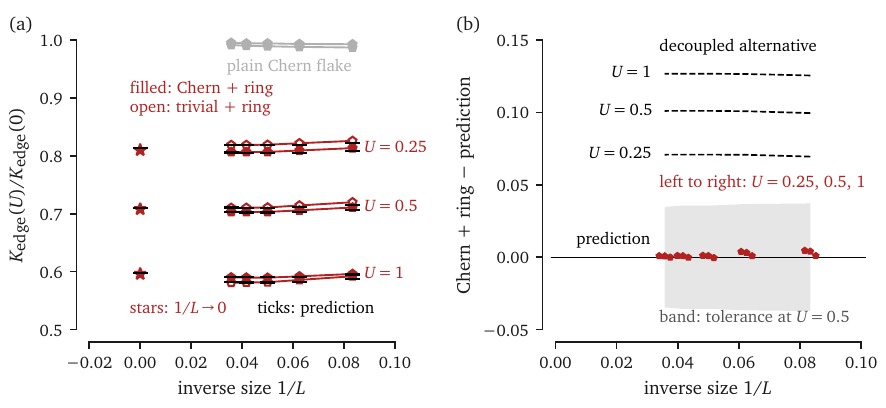}
\caption{\label{fig:coupled}Coupled edge channels in time-dependent Hartree theory at $\lambda=3$, $t_y=0.83$ and $\mu=0.0123\Delta_\mathrm{b}$. (a)~Ratio $K_\mathrm{edge}(U)/K_\mathrm{edge}(0)$ against $1/L$ for the Chern flake with a ring (red, filled), the trivial flake with a ring (red, open) and the plain Chern flake (light grey, shown for $U=0.5$ and 1), with $U$ labelled at each group. Black ticks give Eq.~\eqref{eq:coupled}, fixed for the Chern host before its interacting flakes were run, and $1/\rho$ for the trivial host. Red stars at $1/L=0$ are the extrapolated Chern-host ratios, with the spread of the extrapolation forms as bars. (b)~Chern host minus formula value (red), with the three values of $U$ offset horizontally at each size in the order 0.25, 0.5 and 1 from left to right, and the decoupled alternative minus formula value (black, dashed, labelled by $U$). The solid horizontal line marks the formula value. The band is the tolerance at $U=0.5$.}
\end{figure}

A short-range interaction between counter-propagating edge channels lowers the edge weight by an amount that agrees with a harmonic theory of the coupled channels, while the in-gap Hall weight stays at the net chirality [Fig.~\ref{fig:coupled}].
Take branches $i$ of chirality $\nu_i=\pm1$ on a common boundary, with dipole harmonics obeying $[n_i,n_j^\dagger]=\nu_i\delta_{ij}$, with $\delta_{ij}$ the Kronecker delta, and a Hamiltonian $n^\dagger Vn$ with a positive velocity matrix $V$. The in-gap symmetric weight is then
\begin{equation}
K_\mathrm{edge}=F\,u^{\mathrm{T}}V^{-1/2}\bigl|V^{1/2}NV^{1/2}\bigr|V^{-1/2}u,
\label{eq:coupled}
\end{equation}
where $N=\mathrm{diag}(\nu_i)$, $|\cdot|$ is the matrix absolute value, $u$ is the vector of ones and $F$ is the shape number of Eq.~\eqref{eq:kb}, taken common to all branches. The formula assumes that every branch runs around the same boundary at speeds in a fixed ratio, so that one shape number serves all of them, that the velocity matrix is the same for every harmonic, and that each branch couples to the field through its charge density alone. Appendix~\ref{sec:cb} derives it. The in-gap Hall weight has magnitude $|\sum_i\nu_i|$ for every $V$. For uncoupled branches $V$ is diagonal and Eq.~\eqref{eq:coupled} gives $F$ times the number of branches, and for one chiral branch it gives $F$ at any $V$. This is the edge theory of quantum Hall states with counter-propagating branches~\cite{kane1994randomness,kane1995impurity}, whose velocity matrix depends on the confinement and the interaction~\cite{kane1995impurity}. Diagonalized in the same way, that theory gives each mode of a droplet with several interacting edge modes its own absorption peak with a prefactor set by the interaction~\cite{cano2013microwave}, and Eq.~\eqref{eq:coupled} is the sum of those prefactors. For a non-chiral pair of equal velocities whose intra- and inter-branch couplings are equal, Eq.~\eqref{eq:coupled} gives $K_\mathrm{edge}(U)/K_\mathrm{edge}(0)=1/\rho$, with $\rho$ the factor by which the interaction raises the collective frequency of the pair. This ratio is the Luttinger parameter, which sets the charge fluctuation of a Luttinger liquid~\cite{song2012bipartite}.

We tested Eq.~\eqref{eq:coupled} on the ring flakes of Sec.~\ref{sec:extra} at $t_y=0.83$ and $\mu=0.0123\Delta_\mathrm{b}$, with the time-dependent Hartree theory of that section at $\lambda=3$, for $L=12$ to 28 and $U=0.25$, 0.5 and 1. The Chern host with its ring carries one chiral branch and a counter-propagating pair, and the trivial host carries the pair alone. Its velocity matrix takes the branch velocities from ribbons, the chiral self-coupling from the collective-frequency ratio of the plain Chern flake, the one without a ring, the ring coupling from that of the ring on the trivial host, and a cross coupling equal to the geometric mean of the two, the value for a coupling of rank one. It therefore uses two frequency ratios measured on other flakes and one modelling choice, the cross coupling, and fits nothing to the flake under test. The tolerance was 0.03 plus the amount by which the plain flake misses the ratio of one that Eq.~\eqref{eq:coupled} forces for a single chiral branch, which gives 0.035 to 0.038 at $L=28$. The plain flake itself gives 0.9953 and 0.9915 at $L=28$ and 0.9995 and 0.9994 extrapolated, for $U=0.5$ and 1. The two alternatives were a ratio of one, as for a single chiral branch, and Eq.~\eqref{eq:coupled} with the chiral branch decoupled from the ring.

For the Chern host at $U=0.5$ and 1 the ratio $K_\mathrm{edge}(U)/K_\mathrm{edge}(0)$ is 0.7033 and 0.5903 at $L=28$, against formula values of 0.7025 and 0.5903, and 0.7080 and 0.5961 extrapolated as a quadratic in $1/L$, against 0.7102 and 0.5972, with an extrapolation spread of 0.0069. The decoupled alternative gives 0.8036 and 0.7170 at $L=28$ and is excluded at both, as is a ratio of one. At $U=0.25$ the lattice agrees as well, 0.8068 against 0.8059 at $L=28$, but there the extrapolated formula value lies within twice the tolerance of the decoupled alternative, so this coupling does not separate the two. The in-gap Hall weight extrapolates to magnitude 0.997 at every $U$. The trivial host gives 0.7103 and 0.5811 at $L=28$ against $1/\rho=0.7109$ and 0.5816, with an in-gap Hall weight below $6\times10^{-4}$ in magnitude.
The test fixes the cross coupling more tightly than the branch velocities. At $L=28$ and $U=1$, half and one and a half times the geometric mean move the formula value to 0.6451 and 0.5484, outside the tolerance, while halving or doubling every branch velocity moves it to 0.5617 and 0.6240, inside it. The test therefore supports Eq.~\eqref{eq:coupled} for a coupling of rank one between the branches, the form that a density--density interaction takes, and not for a general velocity matrix.
For the ring flakes the Hartree background is the free density of each site, so that the free state is self-consistent, while the plain flake keeps the uniform background of Appendix~\ref{sec:round2}. A uniform neutralizing background instead drains 18 to 27\,\% of the ring occupancy of the Chern host at $U=1$ for $L=12$ to 20, which changes the branch content, and such flakes are not a test of Eq.~\eqref{eq:coupled}. The transition space is cut at $1.0\Delta_\mathrm{b}$, and moving the cut to $1.5\Delta_\mathrm{b}$ changes the ratio of the Chern host at $L=16$ and $U=1$ by 0.008. The plain flake depends more strongly on the cut. With every particle--hole pair kept (Sec.~\ref{sec:extra}) its ratio at $U=1$ and $L=28$ is 0.964 in place of 0.9915, while its extrapolated ratio stays within 0.001 of one. The agreement of the ring flakes with Eq.~\eqref{eq:coupled} at a given size is therefore an agreement within one transition space. The test covers the random-phase coupling of lattice branches by a screened interaction, and it says nothing about fractional edges or unscreened Coulomb coupling.

\section{Discussion}
\label{sec:discussion}

\subsection{Relation to prior work}

The floor of Eq.~\eqref{eq:floor} joins the edge--bulk cancellation of the dichroic response of Ref.~\cite{unal2024edge}, with its quantization by the many-body Chern number, to the positivity of circular absorption at every frequency, which turns a dichroic sum into a bound on the symmetric weight and underlies the bulk bound of Ref.~\cite{onishi2024bound}.
Reference~\cite{unal2024edge} also shows that the ratio of the two circular edge absorption rates measures the anisotropy of a droplet. In our notation a ratio different from zero means that $K_\mathrm{edge}$ exceeds $|C|$. Its Supplemental Material gives these rates for an arbitrary equipotential boundary, from which Eq.~\eqref{eq:kb} follows (Appendix~\ref{sec:cb}), and adds the channels of concentric circular edges, the circular case of Sec.~\ref{sec:count}.
For a disk the saturated case $K_\mathrm{edge}=|C|$ is familiar from quantum Hall droplets. The projected structure factor of a filled Landau level vanishes~\cite{girvin1986magnetoroton}, and the long-wavelength dipole weight of the guiding centers of a confined droplet can therefore reside only at the edge.
The localization tensor of a bounded sample is known to depend on the sample shape through the screening of the probe field~\cite{resta2011insulating} and on the electrical boundary conditions~\cite{souza2025optical}, and for a Chern flake its boundary term is extensive~\cite{marrazzo2019local}. Eq.~\eqref{eq:floor} bounds that term below by $|C|$ and places it at the edge scale in frequency. Recomputed for their Haldane flake, that term gives an edge weight of 1.165, in line with Eq.~\eqref{eq:kb} (Appendix~\ref{app:mr}). For noninteracting droplets in the lowest Landau level with scale-invariant confinement, Ref.~\cite{oblak2024anisotropic} resolves the edge absorption peaks by the Fourier modes of the boundary deformation, the content of Eq.~\eqref{eq:kb} for that case. The chiral-boson calculation of Ref.~\cite{cano2013microwave} had given such peaks for a droplet of arbitrary shape with one edge velocity, a series of them for each edge mode, and their prefactors for interacting modes, as a way to count the modes and measure their velocities. For droplets in the lowest Landau level the sum of these peaks over frequency is known to be the edge part of the position variance, with the boundary labelled by the canonical angle of the confinement~\cite{moosavi2025quantum}. Secs.~\ref{sec:shape}--\ref{sec:coupled} add to these results the floor of that sum with its Dirichlet form and its condition for saturation, the closed forms, the sums for several and for coupled channels on a boundary that is not circular, and the tests on lattice flakes, which assume no edge theory.

Reference~\cite{onishi2026vacuum} writes the force through the many-body quantum metric of the whole sample and relates that metric to the quantum weight through the structure factor of a translation-invariant system. For a flake the same metric is the static dipole variance of Eq.~\eqref{eq:keff} at $\omega_p\to0$, which already contains the edge.
The force formula therefore holds for a flake as it stands, and what depends on the order of limits is its reading as the bulk quantum weight. For a free-fermion Chern flake at zero temperature, weakly coupled to a circuit slower than the edge, the reading exceeds the bulk quantum weight by an amount that does not vanish with size, because the edge adds at least $|C|$ per unit area and nearly doubles the extrapolated reading of the flakes computed here.
At a fixed coupling a growing flake eventually leaves second order, and the edge share of the shift per unit area then falls (Sec.~\ref{sec:theory}). The adiabatic condition stated in Ref.~\cite{onishi2026vacuum} is, for a flake, exactly the regime in which the edge term is present, so the difference is one of the order of limits. A circuit at fixed $\omega_p$ reads the bulk weight as $L\to\infty$, since $\omega_\mathrm{EMP}$ then falls below $\omega_p$, whereas it reads $K_\mathrm{bulk}+K_\mathrm{edge}$ when $\omega_p\to0$ at fixed $L$.

\subsection{Single-polarization readout}

A device with two orthogonal circuit modes, or one whose field direction can be rotated, reads the trace average of Eq.~\eqref{eq:keff} and therefore the floor. A single field direction reads one diagonal component of the dipole variance, and on the elongated flakes of Sec.~\ref{sec:extra} the short axis reads well below $|C|$, close to $|C|$ times the aspect ratio of the flake.

\subsection{Interactions, temperature and coupling strength}

The numbers of Secs.~\ref{sec:sectorsums}--\ref{sec:routes} hold for free fermions at zero temperature with $\mu$ lying in the bulk gap, coupled through a uniform field at second order in $g$. Secs.~\ref{sec:extra} and \ref{sec:coupled} add screened Hartree flakes at the same order.
Onishi and Fu take the circuit potentials to be approximately uniform within a sample much smaller than the $LC$ circuit~\cite{onishi2026vacuum}, and the uniform-dipole coupling of Eq.~\eqref{eq:hamiltonian} is meant in that sense only.
That coupling is to the bare dipole of the flake in the applied field. Below the edge frequency the charge that the field moves along the edge sets up a depolarizing field of its own. The free-fermion calculation leaves this field out. The time-dependent Hartree calculations of Sec.~\ref{sec:extra} include it. It raises the edge frequency, and on the flakes computed it lowers the in-gap symmetric and Hall weights together and leaves their ratio in place. For an interaction screened at short range by a gate the two weights extrapolate back to their free values, and for the unscreened interaction their extrapolation is unresolved.
The edge term enters the zero-point energy and the force only when three conditions hold at once. The circuit must be slower than the edge, the flake cold compared with the edge scale, and the circuit capacitance large compared with the capacitance added by the edge.
The edge sector contributes zero-point weight only for $k_\mathrm{B}T\ll\hbar\omega_\mathrm{EMP}$, with $k_\mathrm{B}T$ the thermal energy. Otherwise the static edge variance is thermal. The circuit frequency, the readout discussed below, is not bound by this second condition.
The Coulomb interaction sets the velocity of a real edge magnetoplasmon, which is collective and an order of magnitude faster than the drift velocity of the chiral edge state~\cite{emp2024qah}, and adds a logarithm to its dispersion~\cite{volkov1988edge}, which the free-fermion lattice model lacks.
As the flake grows, second-order coupling fails first in the edge sector (Sec.~\ref{sec:theory}), which is exactly where the edge weight enters the reading.

The circuits Onishi and Fu consider operate at microwave energies, far below the electronic gaps of solids~\cite{onishi2026vacuum}, and a measured zero-field edge plasmon of a disk \qty{330}{\micro\meter} across falls in the same microwave band, near \qty{400}{\mega\hertz}~\cite{mahoney2017zero}.
Because the edge scale rises as the inverse perimeter, the flake size decides on which side of the circuit frequency the edge lies.
For the zero-field quantum anomalous Hall insulator Cr-(Bi,Sb)$_2$Te$_3$~\cite{chang2013experimental}, a Chern insulator measured as disks, the fundamental edge magnetoplasmon lies near \qty{0.7}{\giga\hertz} for a disk of radius \qty{125}{\micro\meter} and near \qty{1.5}{\giga\hertz} for one of radius \qty{50}{\micro\meter}, and its dispersion is logarithmic with an edge width of about \qty{1.4}{\micro\meter}~\cite{emp2024qah}.
Fitting that logarithmic form to the five measured disks, with radii from 15 to \qty{125}{\micro\meter}, places the fundamental at \qty{1}{\giga\hertz} for a radius of \qty{69}{\micro\meter} and at \qty{3}{\giga\hertz} for \qty{16}{\micro\meter}. The fundamental reaches \qty{10}{\giga\hertz} only near \qty{2.5}{\micro\meter}, an extrapolation below the measured disks [Fig.~\ref{fig:device}(b), Appendix~\ref{sec:round2}].
A \qty{3}{\giga\hertz} circuit therefore reads the edge term of disks well below \qty{16}{\micro\meter} in radius and the bulk weight of disks well above it, with the leakage $\omega_\mathrm{EMP}/\omega_p$ of Sec.~\ref{sec:model} in between. Because that leakage falls only as the inverse circuit frequency, a bulk reading needs a circuit well above the edge frequency and still below the lowest bulk excitation of the film, which in a disordered magnetic film need not be its nominal exchange gap. Whether such a window exists has to be established for each film.
With the electrons at a cryostat temperature of \qty{20}{\milli\kelvin}~\cite{mahoney2017zero}, which bounds their temperature from below, the thermal factor $\coth(\hbar\omega_\mathrm{EMP}/2k_\mathrm{B}T)$ that multiplies the static variance of an edge mode is 1.46 for the measured \qty{125}{\micro\meter} disk, at its tabulated first resonance of \qty{0.7}{\giga\hertz}, and 1.07 for the \qty{50}{\micro\meter} disk, at \qty{1.4}{\giga\hertz}. It approaches one for the smaller disks, whose edge weight is then zero-point weight, provided the electrons are as cold as the cryostat.

The practical readout of the edge term is the circuit frequency, not the force. For a circuit of capacitance \qty{100}{\femto\farad} whose uniform field spans a gap of twice the disk radius, and with the edge weight at its floor $|C|$, the edge term changes the force on the capacitor plates by $3\times10^{-22}$~N or less, and at the force sensitivity quoted in Ref.~\cite{onishi2026vacuum} a signal-to-noise ratio of one would take at least $2\times10^{6}$~s (Appendix~\ref{sec:coupling}). The force scales as the inverse of the capacitance. At this size and capacitance it is out of reach for the edge term, and for a bulk term of comparable weight. The circuit frequency reads the dipole response of the flake where the force reads its fluctuation. A flake in the gap $d$ of the capacitor adds to it the capacitance $\chi(\omega)/d^2$, with $\chi$ the dipole response along the field (Appendix~\ref{sec:coupling}), and at frequencies well below $\omega_n$ each edge transition contributes to it a capacitance $\Delta C_n$ given by
\begin{equation}
\Delta C_n=\frac{e^2}{h}\,\frac{A}{d^2}\,\frac{s_{a,n}}{\omega_n},
\label{eq:dc}
\end{equation}
where $h$ is the Planck constant and $s_{a,n}=(4\pi/e^2A)|P_{a,n}|^2$ is the weight of the transition along the field direction $a$, so that the $s_{a,n}$ of the edge sector add up to the axis edge weight $K_a$. For a disk with a purely chiral edge, whose lowest edge mode carries the whole edge weight, this is the Hall conductance divided by the edge frequency, times $A/d^2$. Here $d$ is the effective gap, the circuit voltage divided by the field that the edge dipole sees. A disk that fills the gap lies outside the uniform-field limit of Eq.~\eqref{eq:hamiltonian}, so the numbers that follow, for a gap of twice the disk radius, give the scale for the most favourable parallel-plate geometry. They fall as $d^{-2}$ for a wider gap, and a device needs the factor $A/d^2$ from its own electrostatics. It comes to 1.5 to \qty{7.9}{\femto\farad} for radii from 15 to \qty{125}{\micro\meter}, which lowers the frequency of a slower circuit of capacitance \qty{100}{\femto\farad} by 0.8 to 3.7\,\%. A circuit capacitance of \qty{100}{\femto\farad} or more keeps second order accurate to 0.3\,\% at circuit frequencies of 1 to \qty{10}{\giga\hertz} and to 1.9\,\% for a circuit much slower than the edge. These circuit parameters are assumed rather than taken from a device, and the edge frequencies are those of the fitted dispersion. This shift measures the static polarizability of the edge, the edge weight divided by the edge frequency, so the edge weight follows from it once $\omega_\mathrm{EMP}$ is known, for instance from the microwave transmission of the same disk, the measurement that gave the edge-magnetoplasmon dispersion of these films~\cite{emp2024qah}.

A circuit tuned through the edge frequency gives the edge weight directly. The circuit mode and an edge mode repel, and for one edge mode the upper normal frequency $\Omega_{+}$ and the lower normal frequency $\Omega_{-}$ obey $(\Omega_{+}-\Omega_{-})^2=(\omega_n-\omega_p)^2+\omega_n^2\Delta C_n/C_p$ exactly, with $C_p$ the circuit capacitance (Appendix~\ref{sec:coupling}). The smallest splitting $\delta\Omega$, reached at $\omega_p=\omega_n$, therefore gives
\begin{equation}
s_{a,n}=\frac{h}{e^2}\,\frac{C_pd^2}{A}\,\frac{\delta\Omega^2}{\omega_n},
\label{eq:split}
\end{equation}
which takes two measured frequencies and the geometry of the circuit, with neither a force nor a separate measurement of the edge frequency. The weight read in this way is the squared dipole matrix element that also sets the zero-point variance. The avoided crossing therefore reads the edge quantum weight through a response, as the fluctuation--dissipation relation ties the polarization fluctuation of the bulk to its optical absorption~\cite{souza2000polarization}. Summed over the edge modes and averaged over two field directions, the right-hand side is $K_\mathrm{edge}$, which the floor bounds below by $|C|$. For a disk whose edge carries only chiral channels it equals the in-gap Hall magnitude. Where that magnitude is $|C|$, as on the lattice for free fermions and for an interaction screened at short range by a gate (Sec.~\ref{sec:extra}), the avoided crossing measures the Hall conductance without contacts. Counter-propagating channels add to the edge weight (Sec.~\ref{sec:count}), so the splitting then bounds the Hall conductance from above. Its absolute value is as accurate as the geometric factor $A/d^2$, which drops out of the ratio of the weights of two modes of one device. For the measured disks and a circuit capacitance of \qty{100}{\femto\farad} the splitting is 0.12 to 0.28 of the edge frequency. The quality factors of 7 and 19 reported for the disk of radius \qty{125}{\micro\meter}~\cite{emp2024qah} make that splitting 2.0 to 5.3 linewidths, whereas for the disk of radius \qty{15}{\micro\meter}, with a quality factor of 3, it is 0.4 of a linewidth.

The capacitive coupling of a circuit to the edge of a Hall conductor is the principle of the Hall-effect gyrator. For electrodes that cover segments of the edge, Viola and DiVincenzo found an admittance whose poles are the edge-magnetoplasmon frequencies, whose scale is the Hall conductance, and whose low-frequency limit is that of the electrode capacitances alone~\cite{viola2014hall}. Edge magnetoplasmons of a contactless quantum Hall disk have been detected through a nearby transmission line and used to build a circulator~\cite{mahoney2017chip}. Equations~\eqref{eq:dc} and \eqref{eq:split} are the counterpart of that admittance for a uniform field and one lumped circuit mode, in which the weight of each pole is the edge quantum weight of the mode, so that the floor applies to it.

Both readings of the circuit frequency are set by the dipole response of the flake, the expectation value of a commutator, and not by its fluctuation. For harmonic edge modes the response does not depend on temperature, whereas the fluctuation grows as $\coth(\hbar\omega_n/2k_\mathrm{B}T)$. The edge of a lattice flake is a band of fermions and not a set of oscillators, so we tested this on the ellipse and the square at $t_y=0.83$, with the chemical potential held fixed (Appendix~\ref{app:thermal}). Up to $k_\mathrm{B}T=2\hbar\omega_\mathrm{EMP}$ the in-gap response weight, its Hall weight and the static polarizability stay within 1.1\,\% of their values at zero temperature, while the fluctuation grows fourfold, as the bosonic factor gives. The avoided crossing itself follows. With every in-gap pair of the thermal flake coupled to the circuit, the splitting of the doublet stays within 0.7\,\% of its zero-temperature value up to the same temperature, for both flakes, both field directions and splittings of 0.12 and 0.28 of the edge frequency (Appendix~\ref{app:thermal}). At $k_\mathrm{B}T=4\hbar\omega_\mathrm{EMP}$, which is 0.12 to 0.14 of the bulk gap for these flakes, thermally excited bulk carriers change the response weight by 12 to 17\,\% and the splitting by up to 6.7\,\%. The condition $k_\mathrm{B}T\ll\hbar\omega_\mathrm{EMP}$ therefore binds the zero-point energy and the force, and the circuit frequency requires only that the bulk carriers be frozen out. These tests hold the chemical potential fixed and treat free fermions. A fixed electron number, an interacting edge mode and dissipation in the film lie outside them.

For a disk, rotation symmetry lets every in-gap transition of a chiral edge change the angular momentum in the same sense, so the edge sector absorbs one circular polarization only and $K_\mathrm{edge}$ equals the in-gap Hall magnitude, with or without a gate. This assumes that every in-gap channel has the same chirality. A film that also carries counter-propagating or dissipative edge channels absorbs the other polarization as well, and its edge weight then lies above the Hall magnitude. The edge theory with a Coulomb kernel gives the same (Sec.~\ref{sec:coulomb}). The Coulomb interaction then enters Eqs.~\eqref{eq:dc} and \eqref{eq:split} through the edge frequency, which the same measurement gives, and through the in-gap Hall weight itself. On the lattice squares at $U=1$ the in-gap Hall weight at $L=12$ to 28 lies 7.4 to 3.6\,\% below that of the free pairs with a gate and 9.8 to 6.9\,\% below without one (Sec.~\ref{sec:extra}). With a gate the deficit extrapolates to zero, and without one its extrapolation is unresolved. If a lowering survives at the size of a device, the splitting of an ungated disk reads the lowered weight and not the quantized Hall conductance.

With the many-body Chern number in place of $C$, we conjecture that Eq.~\eqref{eq:floor} extends to interacting and fractional Chern flakes whose edge transitions are separated from the bulk by a cut frequency below the bulk gap. The edge--bulk cancellation and its quantization by the many-body Chern number hold for interacting droplets~\cite{unal2024edge}, and the dichroic response of a fractional Chern insulator is fractionally quantized~\cite{repellin2019detecting}. The bulk bound itself holds for the quantum weight of fractional Chern states of twisted bilayer semiconductors computed on periodic clusters~\cite{sfbound2025twisted}. We have not computed a fractional flake, whose edge can carry neutral or counter-propagating modes that such a cut must also separate. In the chiral-boson theory of Appendix~\ref{sec:cb} the commutator of the edge density harmonics does not depend on the interaction, so an interaction that rescales the edge velocity uniformly leaves Eq.~\eqref{eq:kb} unchanged. This is an edge counterpart of the theorem of Kohn, by which a short-range pairwise interaction leaves the cyclotron resonance of a translation-invariant electron gas in a uniform magnetic field at its free frequency~\cite{kohn1961cyclotron}. For a droplet in the lowest Landau level held by a parabolic potential the decoupling of the center of mass of parabolically confined electrons from any pairwise interaction~\cite{maksym1990quantum}, which leaves their dipole absorption independent of the interaction~\cite{brey1989optical}, makes this exact. The edge weight equals the filling factor $\nu$ for any pairwise interaction, which exact diagonalization confirms for Coulomb droplets at $\nu=1/3$ to $10^{-12}$. A non-parabolic confinement adds a positive excess (Appendix~\ref{sec:droplet}).

A circuit mode of one circular polarization, which requires a non-reciprocal circuit, would read the antisymmetric weights instead. In the units of Eq.~\eqref{eq:keff} its helicity difference is $\sum_n a_n\omega_n/(\omega_n+\omega_p)$. It vanishes for a slow circuit, because the edge and bulk dichroic weights of a confined sample cancel~\cite{unal2024edge}. On the lattice flakes its in-gap part falls as $1/(1+\omega_p/\omega_\mathrm{EMP})$ to within 0.011 in every configuration, and at $\hbar\omega_p=0.1\Delta_\mathrm{b}$ the extrapolated reading is 0.82 to 0.92 times $C$ (Appendix~\ref{app:chiral}). The edge has dropped out there, to 0.028 to 0.043 of $|C|$, for the three configurations at $t_y=0.83$. The two at $t_y=0.6$, whose edge frequency at the largest flake is higher, 0.044 and 0.059$\Delta_\mathrm{b}$ against 0.029 to 0.034$\Delta_\mathrm{b}$, still keep 0.076 and 0.132 of $|C|$ in the gap, so the test of a bulk-only chiral reading at that frequency fails for them. This is the frequency-resolved edge sum rule of Ref.~\cite{unal2024edge} read through the zero-point energy rather than through absorption. Vacuum forces set by the Hall conductivity are known between extended quantum Hall and Chern-insulator plates~\cite{tse2012quantized,rodriguez2014repulsive}, and helicity-dependent vacuum shifts have been predicted for a two-dimensional electron gas in a chiral cavity~\cite{kulkarni2026chiral}.

\subsection{Experimental signatures}

Measured as a function of flake size or circuit frequency, the step separates the edge from the bulk contribution to the zero-point shift. In the capacitance that the flake adds to the circuit the separation is sharper still, because Eq.~\eqref{eq:dc} divides each weight by its frequency and the edge frequency is the lowest one.
For the flakes computed here, the inferred quantum weight drops by nearly half across the step. In a real Chern film the bulk quantum weight can far exceed $|C|$~\cite{ghosh2024probing}, and the step is then a smaller fraction of the reading.
The floor turns this step into a sharp prediction. A Chern flake at $k_\mathrm{B}T\ll\hbar\omega_\mathrm{EMP}$, read with both circuit modes averaged and at a coupling weak enough for second order to hold, must give a larger quantum weight well below $\omega_\mathrm{EMP}$ than well above it, and an equal reading on both sides would contradict Eq.~\eqref{eq:floor}. Through Eq.~\eqref{eq:split} the floor is also an inequality between measured frequencies. For edge modes whose avoided crossings with the circuit are resolved one at a time, with splittings $\delta\Omega_n$, the sum of $\delta\Omega_n^2/\omega_n$, averaged over two field directions, must be at least $|C|\,(e^2/h)\,A/C_pd^2$, at any temperature at which the bulk carriers are frozen out. For an ungated flake the bound is the in-gap Hall magnitude, which an unscreened interaction lowers on the lattice flakes at the sizes computed (Sec.~\ref{sec:extra}), and whether it returns to $|C|$ at the size of a device is open.
The size of the step does not by itself identify the Chern number, since a trivial flake with metallic edge states sheds about twice the weight of a Chern flake (Sec.~\ref{sec:extra}). What distinguishes the topological edge term is that it stays at or above $|C|$ for every chemical potential in the bulk gap and carries the in-gap Hall weight $-C$. A trivial edge, by contrast, contributes only while $\mu$ lies inside its own band and carries no Hall weight. A gate sweep of $\mu$ across the bulk gap, or the circular dichroism of the in-gap sector, separates the two. Disorder-bound in-gap states, which carry no Hall weight and need not depend on $\mu$, would weaken the gate test but not the dichroic one. In our disordered ellipse they appear only once the band tails enter the gap, at $W=2\Delta_\mathrm{b}$. There, at $\mu=0.0123\Delta_\mathrm{b}$, they add 0.012 to the in-gap symmetric weight while its Hall weight stays within 0.006 of $-C$, and they add more near the gap edges (Appendix~\ref{sec:round3}).

\section{Conclusion}

A zero-point circuit reads the bulk quantum weight of a finite Chern flake only once it outpaces the edge.
Below the edge-magnetoplasmon frequency the chiral edge adds a weight of at least $|C|$ per unit area, fixed by the edge--bulk cancellation of circular dichroism and the positivity of each transition. The derivation of the floor needs no particular lattice, shape or edge velocity. It needs a bulk Chern number, a chemical potential in the gap, zero temperature and edge transitions that a cut frequency separates from those of the bulk, and the floor then survives as the flake grows. The in-gap Hall weight of $-C$ on which it rests is found on the lattice for free fermions and, in time-dependent Hartree theory, for an interaction screened at short range by a gate, and is left open for an unscreened one. The numerical tests cover three lattice models, the Qi--Wu--Zhang and Haldane models with $|C|=1$ and a two-band model with $|C|=2$. For the Qi--Wu--Zhang flakes computed here, whose bulk weight lies close to its own floor, the extrapolated static reading is nearly twice the bulk value.

Which quantum weight a circuit measures is therefore a question of the order of limits. An infinite flake at fixed circuit frequency gives $K_\mathrm{bulk}$, a slow circuit at fixed flake size gives $K_\mathrm{bulk}+K_\mathrm{edge}$, and in between the edge contribution falls off as $1/(1+\omega_p/\omega_\mathrm{EMP})$. A measurement of quantum geometry on a finite sample has to state which limit it takes before its number can be compared with a band-structure value. The most direct test tunes the circuit frequency or the flake size through the edge-magnetoplasmon frequency of one weakly coupled Chern flake, where the avoided crossing of the two modes gives the edge weight in units of $e^2/h$ [Eq.~\eqref{eq:split}], up to the geometric factor of the capacitor. For the measured Cr-(Bi,Sb)$_2$Te$_3$ disks, a fit to their edge-magnetoplasmon frequencies places a \qty{3}{\giga\hertz} circuit at the edge scale near a radius of \qty{16}{\micro\meter} (Sec.~\ref{sec:discussion}). With the quality factors reported for these disks and the circuit assumed here, the splitting is about two to five linewidths for the largest disk, whose edge mode lies near \qty{0.7}{\giga\hertz}, and less than half a linewidth for the smallest, whose radius is close to that crossing.

The edge term is also a probe in its own right. Its ratio to $|C|$ is the Dirichlet energy of the outline traced at the edge flight time, divided by the area (Sec.~\ref{sec:shape}), which rewrites the edge absorption rates that Ref.~\cite{unal2024edge} gives for an arbitrary equipotential boundary, so a flake read below the edge scale weighs the geometry of its own edge. Each further in-gap channel adds its own shape value while the Hall weight stays at the net chirality (Sec.~\ref{sec:count}), so the two readings together count the non-chiral channels of a clean edge, as the absorption spectrum of a clean droplet does peak by peak~\cite{cano2013microwave}. A screened coupling between counter-propagating channels lowers the edge weight by an amount that agrees with a harmonic theory of the coupled channels (Sec.~\ref{sec:coupled}). Fractional Chern flakes, whose edge dichroism is fixed by the many-body Chern number~\cite{unal2024edge}, are the natural next test of the floor. For lowest-Landau-level droplets the edge weight is bounded below by the filling fraction $\nu$ exactly (Appendix~\ref{sec:droplet}), so we conjecture that the floor carries over with $\nu$ in place of $|C|$. If it does, the zero-field fractional Chern insulators of twisted MoTe$_2$, for which signatures~\cite{cai2023signatures} and a fractionally quantized anomalous Hall effect~\cite{park2023observation} are reported, and of rhombohedral graphene~\cite{lu2024fractional} would carry a fractional edge term in the same circuit.

Several points stay open. With an unscreened interaction the in-gap weights of the lattice flakes lie below those of free fermions at the sizes reached, and whether they recover on larger flakes is open (Sec.~\ref{sec:extra}). A circuit mode of one circular polarization failed its test of a bulk-only reading at the stronger hopping anisotropy (Sec.~\ref{sec:discussion}). The channel counting is untested for edge channels confined to part of the perimeter, and four of the twelve cases with two co-propagating channels are unresolved, among them all those in which the two velocities differ most (Sec.~\ref{sec:count}).

Topology thus puts a floor under a frequency-tunable edge term in capacitive probes of quantum geometry.

\section*{Acknowledgements}
The author gratefully acknowledges the Department of Physics, Indian Institute of Technology Delhi, for computational resources and support.

\begin{locked}The author used Claude (Anthropic) to cross-check numerical values, equations, cross-references and citations against the calculation scripts and primary sources, and verified all of its output.\end{locked}

\paragraph{Data and code availability}
The code and the summary records of the quoted numbers are deposited in the Zenodo record, \href{https://doi.org/10.5281/zenodo.24139452}{doi:10.5281/zenodo.24139452}.

\paragraph{Funding information}
The author received no specific funding for this work.

\begin{appendix}
\numberwithin{equation}{section}

\section{Model and flakes}
\label{app:model}
The flakes are open square-lattice patches of the Qi--Wu--Zhang model~\cite{qi2006qwz} with the Bloch vector of Eq.~\eqref{eq:qwz} and two orbitals per site, the lattice constant, the electron charge $e$ and $\hbar$ being set to one, the area $A$ therefore counting the number of sites. An elliptical flake of linear size $L$ is cut from an $L\times0.75L$ rectangle. The bulk gap $\Delta_\mathrm{b}$ is taken as twice the minimum of $|\mathbf d(\mathbf k)|$ on a dense mesh of wave vectors $\mathbf k$ in the Brillouin zone, and the five chemical potentials, $\mu/\Delta_\mathrm{b}=-0.3877$, $-0.1877$, $0.0123$, $0.2123$ and $0.3877$, are offset from round values to avoid level crossings. The torus bulk weight $K_\mathrm{bulk}$ and the Chern number are obtained from the same $\mathbf d(\mathbf k)$ on a periodic mesh, giving $|C|=1$ for $M=\pm1.37$ and $C=0$ for $M=2.61$. In the sign convention of Eq.~\eqref{eq:weights}, $C=1$ at $M=1.37$ and $C=-1$ at $M=-1.37$. The interior mean of the local Chern marker is $1.000$ and $-1.000$ for the largest squares, $L=48$ and 56, and their bulk Hall weight has the same sign. The torus calculation, with the opposite orientation of the Brillouin zone, records the opposite sign. For every flake and chemical potential, the symmetric weight $s_n$ and the antisymmetric weight $a_n$ of Eq.~\eqref{eq:weights} are evaluated over all particle--hole transitions of the dense eigenbasis, and $K_\mathrm{eff}(\omega_p)$ of Eq.~\eqref{eq:keff} is summed on a logarithmic grid of $\hbar\omega_p/\Delta_\mathrm{b}$. The edge-magnetoplasmon scale $\omega_\mathrm{EMP}$ of each flake is identified with the in-gap transition carrying the largest symmetric weight. The sector sums use a cut at $0.95\Delta_\mathrm{b}$, the in-gap symmetric weight being stored as well at cuts of 0.7, 0.8, 0.9 and $1.0\Delta_\mathrm{b}$ (Fig.~\ref{fig:scans}).

\section{The circuit reading}
\label{app:reading}
The second-order shift of Eq.~\eqref{eq:shift} follows from the Hamiltonian of Eq.~\eqref{eq:hamiltonian} in two terms, the self-energy contributing $(g^2/\hbar\omega_p)\langle0|\hat P_{\hat e}^{\,2}|0\rangle=(g^2/\hbar\omega_p)\sum_{n\neq0}|P_{\hat e,n}|^2$ since $\hat P_{\hat e}$ is the dipole fluctuation, and the linear coupling contributing $-(g^2/\hbar)\sum_{n\neq0}|P_{\hat e,n}|^2/(\omega_n+\omega_p)$ through the intermediate states that carry one circuit quantum. Their sum is Eq.~\eqref{eq:shift}, which, once multiplied by $\hbar\omega_p/g^2$ and averaged over two orthogonal directions, gives Eq.~\eqref{eq:keff}. The reading of Eq.~\eqref{eq:keff} is evaluated on 121 circuit frequencies spaced logarithmically from $10^{-5}\Delta_\mathrm{b}/\hbar$ to $10\Delta_\mathrm{b}/\hbar$ and interpolated linearly in $\ln\omega_p$ between them. The positivity step $|a_n|\leq s_n$ follows from $\bigl||P_{+,n}|^2-|P_{-,n}|^2\bigr|\leq|P_{+,n}|^2+|P_{-,n}|^2$ owing to both terms being nonnegative.

\section{Window estimator and its variants}
\label{app:window}
\begin{figure}[tp]
\includegraphics[width=\textwidth]{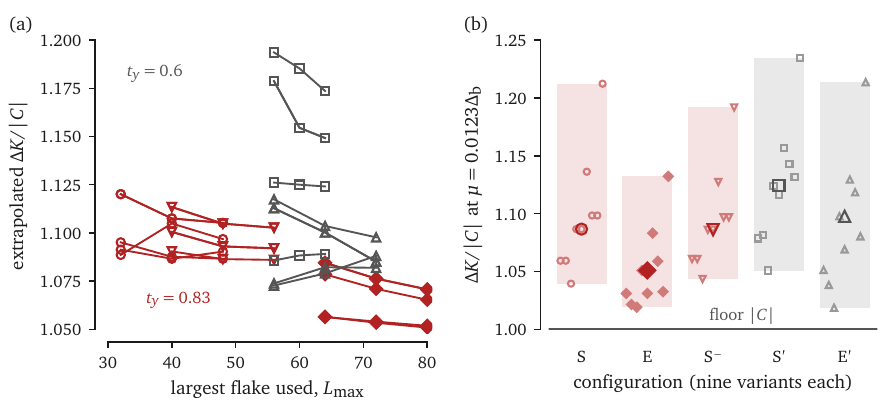}
\caption{\label{fig:conv}Convergence of the window drop. (a)~Extrapolated central $\Delta K/|C|$ against the largest flake used, $L_\mathrm{max}$, for every Chern configuration and chemical potential, with the markers of Fig.~\ref{fig:window}(a). (b)~The nine extrapolation variants at $\mu=0.0123\Delta_\mathrm{b}$ (small markers, with their range as bands) and the central value (large markers) for each configuration. Every value lies above the floor $|C|$, and the threshold 0.9 lies below both plotted ranges.}
\end{figure}

At each flake the window drop $D(L,\kappa)=K_\mathrm{eff}(0)-K_\mathrm{eff}(\omega_p^{*})$ is read at $\omega_p^{*}=\kappa(\omega_\mathrm{EMP}\Delta_\mathrm{b}/\hbar)^{1/2}$ and extrapolated in $x=L^{-1/2}$, since the edge leakage and the bulk loss are both of order $x$, with corrections of order $x^2$.
Nine variants enter the uncertainty at each point, in which each of $\kappa=0.5$, 1 and 2 is fitted as a quadratic in $x$ over all $L\geq24$, as a quadratic over the last five sizes and as a line over the last three.
The central value is $\kappa=1$ with the quadratic over all $L\geq24$.
The spread is the quadrature sum of the variant half-range at the point and the largest change on adding the biggest flake anywhere in the sampled set.
Table~\ref{tab:conv} lists the result at $\mu=0.0123\Delta_\mathrm{b}$ for every configuration, and Fig.~\ref{fig:conv} shows the convergence with the largest flake and the spread of the variants.

\begin{table}[tp]
\caption{\label{tab:conv}Window drop at $\mu=0.0123\Delta_\mathrm{b}$ for each sampled configuration. The columns give the flake shape and parameters, the range of linear sizes $L$, the extrapolated central $\Delta K/|C|$ (for the trivial flake, $\Delta K$), the half-range of the nine extrapolation variants of Appendix~\ref{app:window}, the central value extrapolated with the largest flake removed twice, once, and not at all, and the raw $\kappa=1$ window drop at the largest flake. For the trivial flake only three sizes lie at $L\geq24$, so its fits use all five sizes, and its series of largest flakes has two entries.}
\begin{center}\small\begin{tabular}{>{\raggedright\arraybackslash}p{3.1cm}cccc>{\centering\arraybackslash}p{1.6cm}}
\hline\hline
flake & $L$ & central & half-range & with the three largest $L_\mathrm{max}$ & raw $D(L_\mathrm{max},1)$ \\
\hline
square, $t_y=0.83$, $M=1.37$ & 12--48 & 1.087 & 0.086 & 1.0951, 1.0876, 1.0866 & 0.946 \\
ellipse, $t_y=0.83$, $M=1.37$ & 16--80 & 1.051 & 0.057 & 1.0565, 1.0534, 1.0509 & 0.935 \\
square, $t_y=0.83$, $M=-1.37$ & 16--56 & 1.086 & 0.074 & 1.0904, 1.0864, 1.0860 & 0.959 \\
square, $t_y=0.6$, $M=1.37$ & 16--64 & 1.124 & 0.092 & 1.1261, 1.1250, 1.1241 & 0.896 \\
ellipse, $t_y=0.6$, $M=1.37$ & 16--72 & 1.098 & 0.097 & 1.1174, 1.1036, 1.0977 & 0.867 \\
square, $t_y=0.83$, $M=2.61$ & 12--40 & $-0.0008$ & 0.008 & $-0.0008$, $-0.0008$ & 0.024 \\
\hline\hline
\end{tabular}\end{center}
\end{table}

The prediction is $\Delta K/|C|>0.9$ for the minimum of the extrapolated window drop over the sampled configurations and chemical potentials.
Support requires that adding the biggest flake change the result by less than 0.02 and that the minimum minus twice the spread (defined above) exceed 0.9.
The window of the largest elliptical flake sits at $\hbar\omega_p^{*}=0.156\Delta_\mathrm{b}$, where the edge leakage ratio $\omega_\mathrm{EMP}/\omega_p^{*}$ is also 0.156.
The minimum minus twice the spread is 0.93, which meets this criterion, a criterion on a systematic range and not a confidence interval.
Over all sampled flakes the extrapolated values span 1.05--1.17, the variant half-ranges 0.054--0.117, and the lowest single variant is 0.976. Trivial flakes give $|\Delta K|\leq0.0008$. Larger flakes leave this estimate within its spread (Appendix~\ref{sec:round2}). We add $L=96$ and 112 for the elliptical flake at $t_y=0.83$, 88 and 96 for the ellipse at $t_y=0.6$, 72 and 80 for the square at $t_y=0.6$, and 56 and 64 for the square at $t_y=0.83$. The central values of these configurations shift by $-0.024$ to $+0.011$, and the minimum, at the same flake and $\mu$, moves to 1.046. The raw $\kappa=1$ drop of the largest elliptical flake rises to 0.956, and the raw drops at the new largest sizes are 0.900--0.969.

\begin{figure}[tbp]
\includegraphics[width=\textwidth]{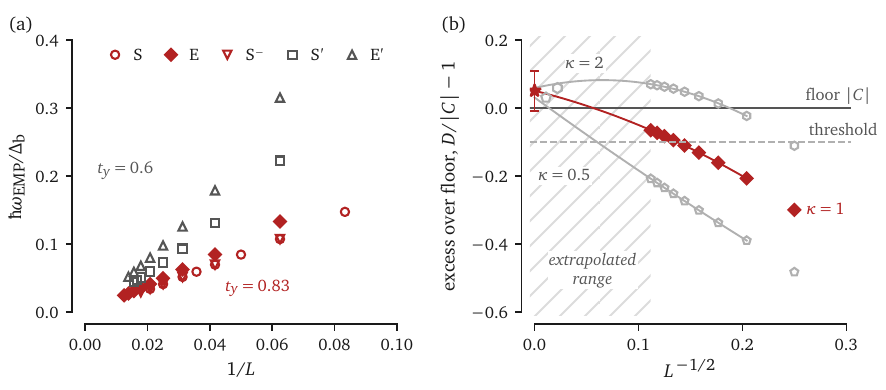}
\caption{\label{fig:window}The edge scale falls with size, and the window drop extrapolates above the floor. (a)~Edge scale $\hbar\omega_\mathrm{EMP}/\Delta_\mathrm{b}$, the in-gap transition of largest symmetric weight, against $1/L$ for the five Chern configurations at $\mu=0.0123\Delta_\mathrm{b}$. (b)~Excess of the window drop over the floor, $D(L,\kappa)/|C|-1$, against $L^{-1/2}$ for the ellipse at $t_y=0.83$ and $\mu=0.0123\Delta_\mathrm{b}$, for window parameters $\kappa=0.5$, 1 (red) and 2. Lines are quadratic fits over $L\geq24$ drawn to $L^{-1/2}=0$, and the star is the extrapolated minimum over the sampled flakes, $\Delta K/|C|-1$, with its spread. The hatched band lies beyond the largest flake ($L=80$). The solid line is the floor $|C|$, and the dashed line is the threshold $\Delta K/|C|=0.9$.}
\end{figure}

\section{Chiral-boson cross-check}
\label{sec:cb}
\begin{figure}[tp]
\includegraphics[width=\textwidth]{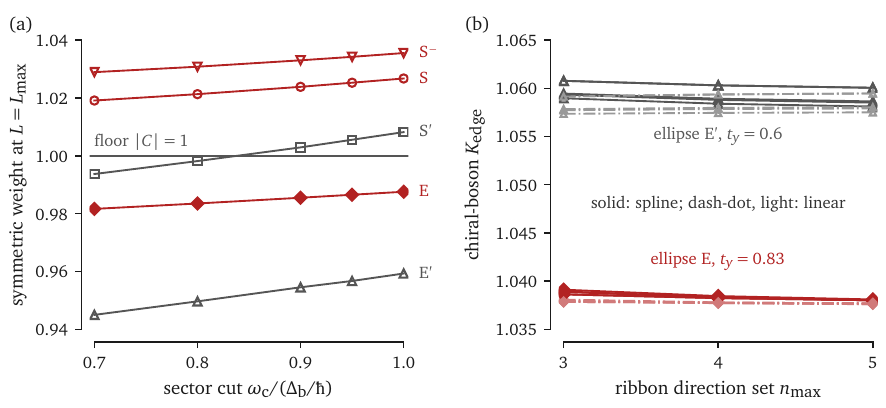}
\caption{\label{fig:scans}Parameter scans. (a)~In-gap symmetric weight of the largest flake of each Chern configuration at $\mu=0.0123\Delta_\mathrm{b}$ against the sector cut $\omega_\mathrm{c}$, in units of $\Delta_\mathrm{b}/\hbar$. The thin line is the floor $|C|=1$. (b)~Analytic chiral-boson edge weight of the elliptical flakes at the five chemical potentials against the ribbon direction-set index $n_\mathrm{max}$ of Appendix~\ref{sec:cb}, with spline (solid) and linear (dash-dot) interpolation of the edge velocity. Red for $t_y=0.83$, dark grey for $t_y=0.6$.}
\end{figure}

The third route computes the edge weight analytically, without diagonalizing the flake, from a chiral-boson edge theory of the kind Wen applied to quantum Hall edges~\cite{wen1990chiral}, through Eq.~\eqref{eq:kb}.
The velocity in each edge direction is the Hellmann--Feynman velocity at $\mu$ of a semi-infinite ribbon cut along a rational lattice direction.
Its numerical parameters are the set of ribbon directions (the primitive lattice directions with integer component $p$ along $x$ and integer component $q$ along $y$ obeying $\max(|p|,|q|)\leq n_\mathrm{max}$ for a direction-set index $n_\mathrm{max}$), the ribbon width and the interpolation of the velocity between directions. Fig.~\ref{fig:scans}(b) shows the direction-set and interpolation dependence for the elliptical flakes. The quoted values use $n_\mathrm{max}=5$, ribbons 30 sites wide sampled at 160 momenta, and a periodic cubic spline of the velocity in the edge direction for the ellipses. For the squares the velocities of the four sides enter directly, from ribbons 60 sites wide sampled at 320 momenta, the corners carry no weight of their own, and ribbons 30 and 45 sites wide change the result by less than $10^{-4}$.
For the elliptical flake at $t_y=0.83$ and $\mu=0.0123\Delta_\mathrm{b}$ it gives 1.0381, and over the sampled flakes the window estimate minus this route runs from $-0.015$ to $+0.069$.

\paragraph*{Several coupled branches.}Eq.~\eqref{eq:coupled} follows from the same commutators. For one harmonic, collect the density harmonics of the branches in a vector $n$ with $[n_i,n_j^\dagger]=\nu_i\delta_{ij}$ and the Hamiltonian $n^\dagger Vn$. The substitution $q=V^{1/2}n$ turns the Hamiltonian into $q^\dagger q$ and the matrix of commutators into $V^{1/2}NV^{1/2}$, which is Hermitian. In the basis that diagonalizes this matrix each component of $q$ is an independent oscillator, an annihilation operator where the eigenvalue is positive and a creation operator where it is negative. The symmetrized ground-state covariance of $q$ is therefore half the matrix absolute value $|V^{1/2}NV^{1/2}|$, and that of $n$ follows by multiplying with $V^{-1/2}$ on both sides. The dipole of the harmonic is the sum of the branch densities times a Fourier coefficient of the boundary, which is the same for every branch when the branches trace the boundary at speeds in a fixed ratio. Summing over the harmonics then gives Eq.~\eqref{eq:coupled}, normalized by the single branch, for which the product of the matrices is one. The Hall weight is the commutator $u^{\mathrm{T}}Nu$ and does not depend on $V$.

\section{Edge variance of a square droplet}
\label{sec:square}
The square value of Sec.~\ref{sec:shape} follows from a Fourier series that can be summed in closed form.
For the potential $U=x$, the edge term of the variance formula of Ref.~\cite{moosavi2025quantum} is a double integral over the canonical angle $\theta$ of the boundary, which at constant edge velocity is linear in arc length,
\begin{equation}
\mathrm{Var}_\mathrm{edge}[x]=\frac{1}{32\pi^2}\oint\!\!\oint\frac{[x(\theta_1)-x(\theta_2)]^2}{\sin^2[(\theta_1-\theta_2)/2]}\,d\theta_1\,d\theta_2=\frac12\sum_{p}|p|\,|x_p|^2 ,
\label{eq:squareedge}
\end{equation}
where $x_p$ are the Fourier coefficients of $x(\theta)$ and the integer $p$ labels them.
A square of area $\pi$ has the side $z(\theta)=(\pi+4\mathrm{i}\theta)/(2\sqrt\pi)$ for $|\theta|\leq\pi/4$, and its other three sides follow by rotation through multiples of $\pi/2$. The Fourier coefficients $z_p$ of $z=x+\mathrm{i}y$ therefore vanish unless $p-1$ is a multiple of four. The magnitudes of the remaining $p$ run once over the odd integers, whose inverse cubes sum to $7\zeta(3)/8$, and since $x$ carries half of each coefficient the edge term is a quarter of the sum of $|p|\,|z_p|^2$, so that
\begin{equation}
|z_p|^2=\frac{32}{\pi^3p^4},\qquad \mathrm{Var}_\mathrm{edge}[x]=\frac{8}{\pi^3}\cdot\frac{7\zeta(3)}{8}=\frac{7\zeta(3)}{\pi^3}.
\label{eq:squaresum}
\end{equation}
For a droplet of area $\pi$ the edge weight of Eq.~\eqref{eq:kb} is four times this variance, which gives $28\zeta(3)/\pi^3$.
Numerically, the sum over modes gives 0.2713774, a direct evaluation of the double integral gives 0.271377, and the same code returns 0.250000 for a disk of area $\pi$.
The mode series printed in the Supplemental Material of Ref.~\cite{moosavi2025quantum}, summed in the same way, gives 0.2683085, which reproduces the quoted edge term 0.2683083. The difference of 0.0031 therefore lies between that printed series and the double integral, not in the numerical evaluation of either.

\section{Per-axis weights, a trivial edge, screening and larger flakes}
\label{sec:round2}
\paragraph*{Per-axis weights.}The flakes of Sec.~\ref{sec:model} were diagonalized again with unchanged Hamiltonian, gap and frequency grid, now storing the in-gap sums of $|P_{x,n}|^2$ and $|P_{y,n}|^2$ separately, together with the per-axis readings $K_{\mathrm{eff},a}(\omega_p)$. The axis weights are extrapolated as the sector sums of Sec.~\ref{sec:results}, quadratically in $1/L$ over the five largest flakes, and the per-axis window drops with the nine variants of Appendix~\ref{app:window}.

\paragraph*{Larger flakes.}The same code, run beyond the recorded sizes, gives $L=96$ and 112 for the elliptical flake at $t_y=0.83$ (7384 sites at $L=112$), 88 and 96 for the ellipse at $t_y=0.6$, 72 and 80 for the square at $t_y=0.6$, and 56 and 64 for the square at $t_y=0.83$. The window estimator of Appendix~\ref{app:window} is applied to the extended series unchanged. Dense diagonalization is kept rather than a sparse or kernel-polynomial route to flakes several hundred sites across, because the edge levels at $\mu$ are spaced by about $\hbar\omega_\mathrm{EMP}$ and any polynomial expansion must resolve that spacing.

\paragraph*{Trivial flake with edge states.}On the trivial flake ($M=2.61$, $t_y=0.83$, square, $L=12$ to 56) one extra orbital is attached to each perimeter site, at the position of that site, with on-site energy zero, hopping $-0.25$ to the extra orbitals of the neighboring perimeter sites, which closes a non-chiral ring, and hybridization $0.3$ with the first orbital of its site. Weights are normalized per site of the host flake. The same ring is attached to the Chern flake ($M=1.37$) as a comparison.

\paragraph*{Screened Hartree flake.}The square Chern flake ($t_y=0.83$, $M=1.37$, $L=12$ to 28) is given a gate-screened interaction between the densities on sites $i$ and $j$ a distance $r_{ij}$ apart, $V_{ij}=U\,e^{-r_{ij}/\lambda}/(r_{ij}^2+1)^{1/2}$ in units of $t_x$ and lattice constants, with screening length $\lambda$, and a neutralizing background. The self-consistent Hartree ground state at fixed electron number is followed by time-dependent Hartree (the random-phase approximation) for the dipole response. In the transition space of all particle--hole pairs below a cut $E_\mathrm{cut}$, the collective modes $\Omega_\alpha$, their dipole matrix elements, and hence their symmetric and antisymmetric weights are computed exactly, and pairs above the cut enter the reading unscreened. With the cut at $1.5\Delta_\mathrm{b}$ the energy-weighted sum $\sum_\alpha\Omega_\alpha s_\alpha$ is not conserved. Independently, the full random-phase response with every pair is evaluated on the imaginary-frequency axis, from which $K_\mathrm{eff}(\omega_p)$ follows by an exact quadrature identity. Doubling its 80 quadrature nodes changes the $U=1$ static reading at $L=8$ by $1.0\times10^{-5}$.

\paragraph*{Device estimate.}The fundamental edge-magnetoplasmon frequencies $f_1$ of the five quantum anomalous Hall disks of Ref.~\cite{emp2024qah} are fitted with the logarithmic disk dispersion $f_1=[v_\mathrm{C}\ln(2R/w+1)+v_\mathrm{D}]/2\pi R$, with $w$ and $v_\mathrm{D}$ fixed at the values of that reference and the Coulomb velocity $v_\mathrm{C}$ free, where $R$ is the disk radius, $w$ the edge width and $v_\mathrm{D}$ the drift velocity. The fit gives $v_\mathrm{C}=8.9\times10^{4}$\,m/s with a largest relative residual of 0.23, the scatter of the measured frequencies, and it predicts \qty{0.49}{\giga\hertz} for the \qty{330}{\micro\meter} disk of Ref.~\cite{mahoney2017zero}, which was not fitted and is measured near \qty{400}{\mega\hertz}. Radii below \qty{15}{\micro\meter} are an extrapolation of the fit. The thermal factor is that of a bosonic edge mode, whose static dipole variance at temperature $T$ is its zero-point value times $\coth(\hbar\omega/2k_\mathrm{B}T)$.

\begin{figure}[tp]
\includegraphics[width=\textwidth]{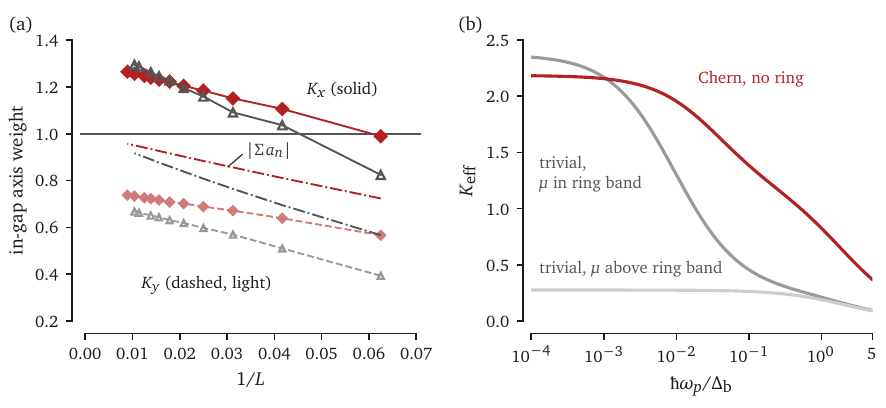}
\caption{\label{fig:round2}Per-axis weights and a trivial edge. (a)~In-gap axis weights $K_x$ (large dark markers, solid lines) and $K_y$ (small light markers, dashed lines) and in-gap Hall magnitude $|\sum a_n|$ (dash-dot) against $1/L$ for the ellipse at $t_y=0.83$ (red filled diamonds) and the ellipse at $t_y=0.6$ (dark grey open up triangles), $\mu=0.0123\Delta_\mathrm{b}$. The thin solid grey line is $|C|=1$. (b)~Circuit reading $K_\mathrm{eff}$ of the trivial flake with a dangling-bond edge ring (Sec.~\ref{sec:extra}) at $L=56$, with $\mu$ inside (darker grey) and above (lighter grey) the ring band, and of the Chern square without a ring at $L=48$ (red).}
\end{figure}

\begin{figure}[tp]
\includegraphics[width=\textwidth]{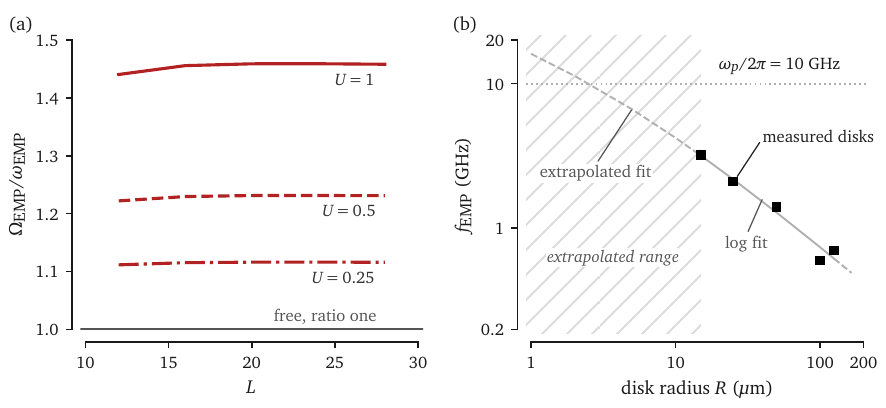}
\caption{\label{fig:device}Screening and a device. (a)~Collective over free edge frequency of the screened Hartree square of Sec.~\ref{sec:extra} against $L$ for $U=0.25$ (dash-dot), 0.5 (dashed) and 1 (solid) at $\lambda=3$. The thin solid line is ratio one, the free case. (b)~Measured fundamental edge-magnetoplasmon frequency of the quantum anomalous Hall disks of Ref.~\cite{emp2024qah} (black filled squares) against radius, the logarithmic fit (grey solid line, dashed where extrapolated below \qty{15}{\micro\meter} and above \qty{125}{\micro\meter}) and the upper end of the circuit band of Sec.~\ref{sec:discussion} (dotted). The hatched band marks the extrapolated range below the smallest measured disk (\qty{15}{\micro\meter}).}
\end{figure}

\section{Exact coupling of the circuit to harmonic edge modes}
\label{sec:coupling}
The second-order shift of Eq.~\eqref{eq:shift} is the leading term of a problem that can be solved exactly when the electronic modes are harmonic, as the chiral-boson edge modes of Appendix~\ref{sec:cb} are. With one circuit mode along the field direction, each electronic mode $n$ written as an oscillator of frequency $\omega_n$ whose coordinate couples through the dipole matrix element $P_n$, and the dipole self-energy term retained, the Hamiltonian of Eq.~\eqref{eq:hamiltonian} is a sum of squares, so it is bounded below and no normal mode goes soft for any coupling $g$. Eliminating the circuit coordinate from the equations of motion and summing the zero-point energies of the normal modes over the imaginary frequency $\xi$ gives
\begin{equation}
\Delta E=\frac{\hbar}{2\pi}\int_0^\infty d\xi\,\ln\!\Bigl[1+\frac{\chi(\mathrm i\xi)}{C_pd^2}\,\frac{\xi^2}{\omega_p^2+\xi^2}\Bigr],
\label{eq:exactshift}
\end{equation}
where $\chi(\mathrm i\xi)=\sum_n2\omega_n|P_n|^2/[\hbar(\omega_n^2+\xi^2)]$ is the dipole response along the field, $C_p$ is the circuit capacitance and $d$ the gap across which its uniform zero-point field $E_\mathrm{zpf}=(\hbar\omega_p/2C_p)^{1/2}/d$ acts, so that $g=E_\mathrm{zpf}$ and $2g^2/\hbar\omega_p=1/C_pd^2$. Expanding the logarithm to first order and integrating returns Eq.~\eqref{eq:shift}, and since $\ln(1+x)\leq x$ the exact shift never exceeds the second-order one. For a single mode the shift is $\Delta E_n=\tfrac12\hbar(\omega_n+\omega_p)[(1+\eta_n)^{1/2}-1]$, with $\eta_n=4g^2|P_n|^2\omega_n/[\hbar^2\omega_p(\omega_n+\omega_p)^2]$, and the relative correction to second order is $-\eta_n/4$ at small $\eta_n$. In the static regime $\eta_n$ tends to $\Delta C_n/C_p$, where $\Delta C_n=2|P_n|^2/(\hbar\omega_nd^2)$ is the capacitance that the polarizable mode adds to the circuit. Without the self-energy term the same model becomes unstable at $\eta_n=1$ in the static regime, which is where second order fails.

\paragraph*{Circuit frequency.}The normal frequencies $\Omega$ of the circuit coupled to the harmonic modes obey $\Omega^2[1+\chi(\Omega)/C_pd^2]=\omega_p^2$, with $\chi(\Omega)=\sum_n2\omega_n|P_n|^2/[\hbar(\omega_n^2-\Omega^2)]$ the dipole response at real frequency. This is the resonance of a circuit of capacitance $C_p+\chi(\Omega)/d^2$, and for $\Omega\ll\omega_n$ each mode adds the capacitance $\Delta C_n$ of Eq.~\eqref{eq:dc}. With one electronic mode the two roots obey $\Omega_{+}\Omega_{-}=\omega_p\omega_n$ and $(\Omega_{+}+\Omega_{-})^2=(\omega_n+\omega_p)^2+\omega_n^2\Delta C_n/C_p$, so that $(\Omega_{+}-\Omega_{-})^2=(\omega_n-\omega_p)^2+\omega_n^2\Delta C_n/C_p$, the relation behind Eq.~\eqref{eq:split}. At fixed $C_p$ the last term does not depend on $\omega_p$, and the smallest splitting is $\omega_n(\Delta C_n/C_p)^{1/2}$. The response $\chi$ is the expectation value of a commutator. For harmonic modes it does not depend on temperature, unlike the dipole variance, which carries the factor $\coth(\hbar\omega_n/2k_\mathrm{B}T)$ (Appendix~\ref{app:thermal}).

 In the edge sector at fixed coupling, $|P_n|\propto L$ and $\omega_n\propto1/L$ make $\eta_n$ grow as $L^3$. The computed edge share of the shift per unit area then falls with exponent $-1.50$ in $L$, so $K_\mathrm{edge}$ is the weak-coupling coefficient of the reading.

For the quantum anomalous Hall disks of the device estimate, with the edge weight at its floor $|C|$ and the disk spanning the gap ($d=2R$), the edge adds 1.5 to \qty{7.9}{\femto\farad} to the circuit for radii from 15 to \qty{125}{\micro\meter}. Over 90 assumed circuits, with $\omega_p/2\pi=1$, 3 and \qty{10}{\giga\hertz}, $C_p=10$, 100 and \qty{1000}{\femto\farad} and $d=2R$ or $10R$, the largest $\eta_n$ is 0.12, an error of second order of 2.8\,\%, and five circuits exceed 1\,\%, all at the smallest capacitance. For $C_p\geq\qty{100}{\femto\farad}$ or $d=10R$ the error is at most 0.3\,\%. The calculation treats one linearly polarized circuit mode at zero temperature. Two orthogonal modes coupled simultaneously are not treated, and lattice edge transitions are harmonic only approximately.

\paragraph*{Size of the force.}For the circuit assumed in Sec.~\ref{sec:discussion}, with $C_p=\qty{100}{\femto\farad}$, parallel plates $2R$ apart and the edge weight at its floor, the second-order shift of Eq.~\eqref{eq:shift} in the static regime is $(eR/2)^2/(2C_pd^2)$ per field direction, which at $d=2R$ is $5.0\times10^{-8}$~eV for every measured disk. At fixed plate area this shift varies as $1/d$, so the edge term changes the force on the plates by $3\times10^{-23}$ to $3\times10^{-22}$~N for radii from 125 to \qty{15}{\micro\meter}, a fraction $3\times10^{-5}$ to $5\times10^{-7}$ of the Casimir force between ideal plates of the same capacitance. At the force sensitivity of $390$~zN\,Hz$^{-1/2}$ quoted in Ref.~\cite{onishi2026vacuum} a signal-to-noise ratio of one would take at least $2\times10^{6}$~s. With the measured frequencies the same edge weight adds 1.5 to 8.1 per cent to the circuit capacitance, so for a single disk the circuit frequency, not the force, is the practical readout.

\section{Common size set, larger flakes, disks, disorder and unscreened interaction}
\label{sec:round3}
None of the tests in this appendix changes a result reported for the flakes of the main series.

\paragraph*{Common size set.}The sector sums of Sec.~\ref{sec:sectorsums} are extrapolated over the five largest flakes of each configuration, including the larger flakes of Appendix~\ref{sec:round2} but not the $t_y=0.6$ extension below. Over the pilot sizes, the sizes of Table~\ref{tab:conv} and the five largest flakes, in that order, the extrapolated in-gap Hall magnitude spans 0.998--1.009, 0.998--1.012 and 0.998--1.007, the in-gap symmetric weight 1.03--1.14, 1.01--1.13 and 1.04--1.13, and the largest bulk-sector residual is 0.038, 0.017 and 0.0085, over all configurations and chemical potentials.

\paragraph*{Larger flakes at $t_y=0.6$.}To test whether the excess of the in-gap symmetric weight over $|C|$ settles at $t_y=0.6$, the square is extended to $L=88$, 96 and 104 (10816 sites) and the ellipse to $L=104$, 112, 128 and 144 (12212 sites). The excess is extrapolated quadratically in $1/L$ over five consecutive sizes, and it was declared settled if the last three such windows differ consecutively by at most 0.02 and those differences do not change sign. With windows ending at $L=96$ and 128 the rule settled the excess at $\mu=0.0123\Delta_\mathrm{b}$ only. The sizes $L=104$ and 144 test whether this settlement survives a further window. It does not, since at $\mu=0.0123\Delta_\mathrm{b}$ the last three windows give 0.102, 0.101 and 0.104 for the square and 0.056, 0.067 and 0.051 for the ellipse, so the differences, though below 0.02, change sign for both shapes. Of the other chemical potentials the rule is met only at $\mu=-0.1877\Delta_\mathrm{b}$ for both shapes and at $0.2123\Delta_\mathrm{b}$ for the ellipse, and window-to-window changes reach 0.043. At every chemical potential the in-gap symmetric weight of the largest flake exceeds $|C|$, 1.043--1.058 for the square at $L=104$ and 1.007--1.016 for the ellipse at $L=144$, while the in-gap Hall magnitude extrapolated over the last window lies within 0.004 of one.

\paragraph*{Lattice disks.}Disks of diameter $L=16$ to 96 are cut from the model of Eq.~\eqref{eq:qwz} with $t_x=t_y=1$ and $M=1.37$, with $b=0$, for which the flake is symmetric under fourfold rotation, and with $b=0.11$, which breaks that symmetry. The chiral-boson route of Appendix~\ref{sec:cb}, with the ribbon velocities of the same parameters, returns one to $2\times10^{-10}$ for a disk at constant velocity. Extrapolated quadratically in $1/L$ over the five largest disks, with the spread taken as the half-range over five fit variants, the in-gap symmetric weight is 1.006--1.012 at $b=0$ and 1.010--1.012 at $b=0.11$ over the five chemical potentials, with spreads of 0.011--0.023, and the in-gap Hall magnitude is 1.000--1.004, and the chiral-boson values are 1.001--1.002. Neither the disk value minus one nor the disk value minus the chiral-boson value exceeds twice its spread. The spread is larger than for the ellipse because the disk series approaches $|C|$ from below with stronger curvature, its in-gap symmetric weight being 0.974--0.975 at $L=96$, so the predicted excess of order $10^{-3}$ is not resolved.

\paragraph*{Anderson disorder.}A random on-site energy, drawn uniformly from $[-W/2,W/2]$ and equal on the two orbitals of a site, is added to the ellipse ($L=16$ to 64) and the square ($L=12$ to 48) at $t_y=0.83$, with 16 realizations per size and the seed of every realization stored. The sector cut stays at $0.95\Delta_\mathrm{b}$ of the clean flake. Table~\ref{tab:disorder} lists the extrapolated disorder averages at $\mu=0.0123\Delta_\mathrm{b}$ against the clean values on the same sizes and estimator, whose ellipse value differs from that of Sec.~\ref{sec:sectorsums} because the series stops at $L=64$. For $W\leq\Delta_\mathrm{b}$ the in-gap Hall weight stays at $-C$ at all five chemical potentials, within three standard errors plus the extrapolation spread of its clean value, and the in-gap symmetric weight stays above $|C|$, at least 1.037. It is within 0.003 of its clean value at $\mu=0.0123\Delta_\mathrm{b}$ but falls by up to 0.013, 3.5 to 4.3 standard errors, for the ellipse at $W=\Delta_\mathrm{b}$ and $\mu=0.2123$ and $0.3877\Delta_\mathrm{b}$, where the window drop falls by 0.03 to 0.04. At $W=2\Delta_\mathrm{b}$, when band-tail states enter the clean-gap window, the ellipse gains $0.012\pm0.003$ of in-gap symmetric weight while its Hall magnitude falls by 0.004. At $W=4\Delta_\mathrm{b}$ the in-gap Hall magnitude falls to $0.867\pm0.013$, so the Hall test can fail. Its finite-size value is still rising, from 0.70 at $L=16$ to 0.82 at $L=64$, so whether the mobility gap has closed or the localization length exceeds the flake is not resolved, and the dominant in-gap transition there is a tail-state transition rather than an edge mode.

\begin{table}[tp]
\caption{\label{tab:disorder}Extrapolated in-gap symmetric weight $K_\mathrm{edge}$, in-gap Hall weight and $\kappa=1$ window drop $D$ under Anderson disorder at $\mu=0.0123\Delta_\mathrm{b}$, $t_y=0.83$, $M=1.37$, averaged over 16 realizations, with the standard error of the average.}
\begin{center}\begin{tabular}{lccc}
\hline\hline
flake, $W/\Delta_\mathrm{b}$ & $K_\mathrm{edge}$ & Hall & $D$ \\
\hline
ellipse, 0 & 1.0407 & $-0.9990$ & 1.055 \\
ellipse, 0.5 & $1.0405\pm0.0004$ & $-0.9992\pm0.0002$ & $1.054\pm0.001$ \\
ellipse, 1 & $1.0426\pm0.0013$ & $-0.9986\pm0.0007$ & $1.055\pm0.004$ \\
ellipse, 2 & $1.0531\pm0.0032$ & $-0.9949\pm0.0022$ & $1.069\pm0.010$ \\
square, 0 & 1.0901 & $-0.9988$ & 1.087 \\
square, 0.5 & $1.0874\pm0.0023$ & $-0.9996\pm0.0003$ & $1.083\pm0.007$ \\
square, 1 & $1.0896\pm0.0043$ & $-0.9994\pm0.0012$ & $1.088\pm0.012$ \\
\hline\hline
\end{tabular}\end{center}
\end{table}

\paragraph*{Unscreened interaction.}The screened Hartree square of Appendix~\ref{sec:round2} is repeated with $\lambda=L/2$, $L$, $2L$ and $\lambda\to\infty$, at $U=1$ for all four and at $U=0.5$ for $L/2$ and $\infty$, for $L=12$ to 28. At $U=1$ the collective over free edge frequency rises from 1.56 to 1.85 between $L=12$ and 28 at $\lambda=L/2$ and from 1.64 to 1.96 at $\lambda\to\infty$. At $L=24$ it fits $1.194+0.245\ln\lambda$ for $\lambda\leq12$ and saturates by $\lambda\approx2L$. The dominant edge mode stays circular to $1.2\times10^{-4}$ at every unscreened point. Extrapolated as at $\lambda=3$, the in-gap symmetric weight falls below its free value by 0.023--0.025 at $U=1$, over the four screening lengths, and by 0.013--0.014 at $U=0.5$, together with a fall of the in-gap Hall magnitude to 0.983--0.985 at $U=1$, so the ratio of the two barely moves, 1.077 against 1.080. Moving the cut from 1.5 to $2.5\Delta_\mathrm{b}$ at $\lambda\to\infty$ lowers both weights by 0.025--0.026 at $L=12$ and 16 and is not converged there, while with only the in-gap pairs the Hall weight equals the free one to $9\times10^{-4}$. The shift therefore comes from mixing with pairs above the gap, which the truncation does not resolve. With every pair included, on the imaginary axis, the window drop at the screened edge scale extrapolates to 1.100 for $L=12$ to 24 and 1.099 for $L=12$ to 28 at $\lambda\to\infty$, against 1.098 without interaction and 1.142 at $\lambda=3$, while a linear form in $L^{-1/2}$ over $L=12$ to 24 gives 1.118 at $\lambda\to\infty$ and 1.198 without interaction. The unscreened interaction does not move the window drop outside the spread between extrapolation forms.

\paragraph*{Every particle--hole pair.}The cut of the transition space is removed by applying the density response of the time-dependent Hartree theory, with every pair, to the dipole potentials at complex frequencies on a circle through zero and $0.95\Delta_\mathrm{b}$, and by integrating around the circle with 96 nodes. The integral gives the in-gap sums through a soft cut at that frequency. The same soft cut is applied to the pairs of the mean field without their residual interaction, so the two are compared on one estimator. The response is applied to a potential without being built, and the equation of the random-phase approximation is solved iteratively. Three criteria are applied, on squares of $L=12$ to 28 for $U=1$ and 0.5 without a gate and for $U=1$ at $\lambda=3$. Each extrapolation is a quadratic in $1/L$ through the five sizes, and its spread is half the range between that fit, a quadratic through the four largest sizes and a line through the three largest. The in-gap Hall magnitude should extrapolate to one within the larger of 0.02 and the spread. The ratio of symmetric to Hall weight, divided by that of the free pairs of the same mean field, should lie at every size between 0.982, its value with the in-gap pairs alone, and one, as it does if the pairs of the bulk only screen the edge. The ratio of the two extrapolated weights should lie within the larger of 0.01 and its spread of that of the free flake. On a small flake the contour sums agree with the sums over the modes of the full matrix of the theory to $4\times10^{-13}$. The first criterion is met at $\lambda=3$, with 1.003, and at $U=0.5$ without a gate, with 0.987, and it is missed at $U=1$ without a gate, where the three fits give 0.971, 0.974 and 0.949. The second is met without a gate, with 0.9955--0.9960 at $U=1$ and 0.9964--0.9969 at $U=0.5$, and is missed at $\lambda=3$, where the ratio exceeds one at the three smallest sizes, by up to 0.0021. The third is met in all three series. The ratio of the two extrapolated weights is 1.079 at $U=1$ and 1.078 at $U=0.5$ without a gate and 1.081 at $\lambda=3$, against 1.079 for the free flake. The in-gap symmetric weight extrapolates to 1.048 and 1.064 without a gate at $U=1$ and 0.5 and to 1.084 at $\lambda=3$, against 1.083 for the free flake. The deficit of the truncated calculation is therefore not an artefact of its cut at these sizes. At $U=1$ the symmetric weight lies 10.2\,\% below that of the free pairs at $L=12$ and 7.3\,\% below at $L=28$ without a gate, against 7.2 and 3.6\,\% at $\lambda=3$, and a soft cut at $0.8\Delta_\mathrm{b}$ on the two smallest flakes moves these deficits by at most 0.14 percentage points. Without a gate the fit carries the Hall magnitude from 0.817 at $L=28$ to 0.971. A quadratic in $1/L$ fitted to the Hall deficit leaves 2.6\,\% on large flakes, and a form that leaves nothing, inversely proportional to the size plus a constant, fits the five sizes equally well, with a root-mean-square residual of $1.6\times10^{-4}$ in both cases. Five sizes therefore do not decide whether the deficit vanishes on larger flakes. We treat the separate extrapolations without a gate as unresolved and claim only that the interaction leaves the ratio of the two weights in place.

\section{Confined Landau-level droplets}
\label{sec:droplet}
For a lowest-Landau-level droplet with electron number $N$ and magnetic length $l$, the dipole couples only to the guiding centers. With $b_i$ the guiding-center lowering operator of electron $i$ and their sum $B=\sum_ib_i$, so that $[B,B^\dagger]=N$, and with the droplet area taken from the bulk density, $A=2\pi l^2N/\nu$, the intra-level symmetric and antisymmetric weights $K_\mathrm{edge}$ and $A_\mathrm{edge}$ are
\begin{equation}
K_\mathrm{edge}=\nu\left(1+\frac{2\langle B^\dagger B\rangle}{N}\right),\qquad A_\mathrm{edge}=\frac{\nu}{N}\left(\langle BB^\dagger\rangle-\langle B^\dagger B\rangle\right)=\nu,
\label{eq:llldroplet}
\end{equation}
where $\langle\cdot\rangle$ is the ground-state expectation value, the first form assumes $\langle B\rangle=0$, as for any state of fixed angular momentum, the sign of $A_\mathrm{edge}$ is that of a field orientation in which $P_+\propto B^\dagger$, and $A$ is a normalization rather than the area of a compressed droplet. The antisymmetric weight is fixed by the commutator for every state of the level, the lowest-Landau-level form of the edge sum rule, while the symmetric weight saturates the floor exactly when $B$ annihilates the ground state. Under a parabolic confinement the center of mass decouples from any pairwise interaction~\cite{maksym1990quantum}. The confinement then shifts $B$ by a constant energy, so the ground state has $B|0\rangle=0$ and $K_\mathrm{edge}=\nu$.

\paragraph*{Exact diagonalization.}The droplet Hamiltonian is diagonalized with Haldane pseudopotentials for the Coulomb interaction and for the short-range interaction $V_1$, for $N=3$ to 6 electrons at $\nu=1$ and 1/3. At a filling of one third the droplet is in the quantum-fluid state that Laughlin described with a variational wave function~\cite{laughlin1983anomalous}. With a parabolic confinement the edge weight of the $\nu=1/3$ droplet equals $\nu$ to $10^{-15}$ for $V_1$ and, once the orbital cutoff is converged, to $10^{-12}$ for the Coulomb interaction at $N=4$ to 6. At $\nu=1$ the parabolic runs did not bracket the ground state in angular momentum and are not used. A quartic term added to the confinement leaves the maximum-density droplet at $\nu=1$ exact, because it is the only state of its angular momentum, but raises the edge weight of the $V_1$ Laughlin droplet above $\nu$ by $2.0\times10^{-4}$ to $5.1\times10^{-4}$, with up to $5\times10^{-5}$ of the weight lost to the orbital cutoff, so the saturation is then not exact. A quartic potential is not constant within a sector of fixed angular momentum, and it mixes the Laughlin state with others. With the Coulomb interaction at $\nu=1/3$ the excess is $1.4\times10^{-3}$ to $5.1\times10^{-3}$ and $8.3\times10^{-3}$ to 0.024 for the two quartic strengths, and it is larger at $N=6$ than at $N=4$ for both strengths, against the decrease with $N$ we had predicted.

\paragraph*{Elliptic traps.}An elliptic harmonic confinement with squeezing parameter $\beta$, $\propto e^{2\beta}x^2+e^{-2\beta}y^2$, still decouples the center of mass, which is then squeezed, so $K_\mathrm{edge}=\nu\cosh2\beta=\nu(r+1/r)/2$ for a droplet of aspect ratio $r=e^{-2\beta}$. This is the symmetric combination of the two circular edge rates, whose ratio Ref.~\cite{unal2024edge} uses as a measure of the anisotropy of a droplet and which Eq.~\eqref{eq:llldroplet} puts at $\tanh^2\beta$. Coulomb droplets of $N=3$ and 4 electrons come within $4\times10^{-9}$ of it at $\beta=0.1$ and $2\times10^{-5}$ at $\beta=0.2$, but enlarging the orbital cutoff still moves them by up to $2\times10^{-6}$ and $8.1\times10^{-4}$, so the agreement is not converged. Eq.~\eqref{eq:kb} for an ellipse traversed with the harmonic-trap velocity gives $(r+1/r)/2$ to $3\times10^{-10}$ for $r$ from 0.5 to 1, whereas at constant velocity it gives 1.024 at $r=0.75$ against 1.042. The excess of $K_\mathrm{edge}$ over $|C|$ therefore depends on how the edge velocity varies around the boundary as well as on its outline.

\section{Chiral circuit reading}
\label{app:chiral}
The chiral reading $Q(\omega_p)=\sum_na_n\omega_n/(\omega_n+\omega_p)$ is computed on the flakes of Sec.~\ref{sec:sectorsums} with $L\leq64$, with the same code, sector cut $0.95\Delta_\mathrm{b}$ and $\mu=0.0123\Delta_\mathrm{b}$, and is split into in-gap and bulk parts as the symmetric reading is. At each of $\hbar\omega_p/\Delta_\mathrm{b}=0.1$, 0.2, 0.3 and 0.5 the reading is extrapolated quadratically in $1/L$ over $L\geq24$, and Table~\ref{tab:chiral} lists the result at $0.1\Delta_\mathrm{b}$.
The criteria were a reading within 0.85 to 1.05 of $C$ and an in-gap part of at most $0.05|C|$, both for every configuration. Both hold for the three configurations at $t_y=0.83$, and both fail for the ellipse at $t_y=0.6$, while the square at $t_y=0.6$ meets the first and fails the second. The in-gap part falls with circuit frequency in every configuration, to at most 0.009 of $|C|$ at $0.5\Delta_\mathrm{b}$. The bulk part, 0.94 to 0.96 of $C$ at $0.1\Delta_\mathrm{b}$, lies below $C$ because the factor $\omega_n/(\omega_n+\omega_p)$ reduces transitions near the gap. The extrapolation is large, since at the largest flake the reading at $0.1\Delta_\mathrm{b}$ is 0.52 to 0.69 of $C$.
At the largest flake the in-gap reading, relative to its static value, follows $1/(1+\omega_p/\omega_\mathrm{EMP})$ to within 0.011 at $\omega_p/\omega_\mathrm{EMP}=0.3$, 1, 3 and 10 in every configuration, with $\omega_\mathrm{EMP}$ taken as the frequency of the largest in-gap symmetric weight. This law follows largely from the in-gap weight lying near a single frequency, so it is a consistency check rather than independent evidence.

\begin{table}[tp]
\caption{\label{tab:chiral}Chiral reading at $\hbar\omega_p=0.1\Delta_\mathrm{b}$ and $\mu=0.0123\Delta_\mathrm{b}$, in units of $C$. The columns give the flake, the largest linear size $L_\mathrm{max}$, the extrapolated reading $Q_\infty$, its in-gap part, its bulk part (the two add to $Q_\infty$), and the reading at $L_\mathrm{max}$.}
\begin{center}\small\begin{tabular}{>{\raggedright\arraybackslash}p{4.2cm}ccccc}
\hline\hline
flake & $L_\mathrm{max}$ & $Q_\infty/C$ & in-gap & bulk & $Q(L_\mathrm{max})/C$ \\
\hline
square, $t_y=0.83$, $M=1.37$ & 48 & 0.911 & $-0.033$ & 0.943 & 0.652 \\
ellipse, $t_y=0.83$, $M=1.37$ & 64 & 0.900 & $-0.043$ & 0.943 & 0.661 \\
square, $t_y=0.83$, $M=-1.37$ & 56 & 0.916 & $-0.028$ & 0.944 & 0.685 \\
square, $t_y=0.6$, $M=1.37$ & 64 & 0.880 & $-0.076$ & 0.957 & 0.593 \\
ellipse, $t_y=0.6$, $M=1.37$ & 64 & 0.822 & $-0.132$ & 0.955 & 0.516 \\
\hline\hline
\end{tabular}\end{center}
\end{table}

\section{The Haldane flake of Marrazzo and Resta}
\label{app:mr}
We recomputed the Chern flake of Ref.~\cite{marrazzo2019local} with the parameters of its Supplemental Material, the Haldane model with first-neighbor hopping 1, second-neighbor hopping $1/3$ with phase $\pi/4$ and sublattice energies $\pm1/3$, on rectangular honeycomb flakes of $N=2p(2p+1)$ sites with zigzag and armchair edges, $p=5$ to 45.
The trace plotted in their Fig.~4 follows Eq.~(1) of their Supplemental Material and is twice the second cumulant of the dipole per area, so it equals $K/\pi$.
With exactly $N/2$ electrons all 21 points of their figure are reproduced to within 0.0006.
At zero chemical potential the in-gap symmetric weight extrapolates to 1.165, with 1.157--1.173 over the extrapolation variants, and the in-gap Hall magnitude to 1.002.
Eq.~\eqref{eq:kb} with the ribbon velocities of the four sides gives 1.1692.
Constant velocity gives 1.1695, so this flake tests the shape number and not the velocity dependence.

\section{A Chern band with $|C|=2$}
\label{app:c2}
The two-orbital model of Sec.~\ref{sec:count} has the Bloch vector
\begin{equation}
\mathbf d=\bigl(a(\cos k_x-\cos k_y)+b,\;2c\sin k_x\sin k_y,\;M-t_x\cos k_x-t_y\cos k_y\bigr),
\label{eq:c2}
\end{equation}
with $a$ and $c$ the amplitudes of the two inter-orbital hoppings, to first neighbours along the axes and to second neighbours along the diagonals. The parameters $t_x=1$, $t_y=0.83$ and $b=0.11$ are those of the main series, and $M=1$, $a=1$ and $c=0.75$ give a lower band of Chern number magnitude two, a bulk gap of at least 0.9 and two edge branches of the same chirality on each of the four sides at all five chemical potentials. The bulk gap is 1.018 and the torus bulk weight 2.153, above its own floor of two.
On a side along $x$ the two branches cross a given energy at opposite momenta, with the same velocity at the centre of the gap. At $\mu=0.0123\Delta_\mathrm{b}$ their velocities are 1.229 and 1.215 on the sides along $x$ and 1.038 and 1.019 on the sides along $y$, and at $\mu=\pm0.3877\Delta_\mathrm{b}$ they are 1.375 and 0.887, and 1.234 and 0.532. The ribbon spectra at $\mu$ and $-\mu$ are mirror images, so the two outermost chemical potentials are one test and each shape has four independent ones.
The hypothesis of Sec.~\ref{sec:count} assigns Eq.~\eqref{eq:kb} to each branch at its own velocities. Two co-propagating branches can exchange at a corner, and the prediction is taken as the range over the sixteen ways in which the faster branch of one side can continue into the faster or the slower branch of the next. An odd number of exchanges joins the two branches into one loop that winds twice. When both branches have the same velocity on every side the sixteen values coincide. The prediction of an in-gap Hall weight of magnitude two and this range are judged by the rule of Sec.~\ref{sec:count}, against three alternatives, namely one branch alone, the square value at any shape, and an edge weight at its floor. One branch alone would violate Eq.~\eqref{eq:floor} once the in-gap Hall magnitude is two, so the informative alternatives are the square value and the floor.
Squares of $L=28$ to 88 and 2:1 rectangles in both orientations, with up to 8192 sites, were diagonalized and extrapolated with the estimator of Sec.~\ref{sec:count}, over the three largest windows of five consecutive sizes. A spread between the windows above 0.05 counts as unresolved, and the tolerance is the larger of 0.03 and the spread. On the seven smallest sizes of each series, with up to 5184 sites, seven of the twelve cases are resolved and agree with the prediction and five are unresolved. The two largest sizes resolve two of those five and leave unresolved one case that the seven sizes resolve, the rectangle with the long side along $y$ at $\mu=\pm0.3877\Delta_\mathrm{b}$, whose spread between windows grows from 0.005 to 0.066. A small spread between three windows therefore does not by itself establish convergence.
The same hoppings at $M=2.61$, where the band is trivial, give no in-gap weight at all. Extrapolated over the same windows, the bulk-sector symmetric weight returns the torus bulk weight to within 0.0021 in every case, and over all computed sizes the smallest ratio of the in-gap symmetric weight to the in-gap Hall magnitude is 1.108. Table~\ref{tab:c2} lists every case.

\paragraph*{Level-averaged sums.}The unresolved cases pointed to the discreteness of the edge levels. At a fixed chemical potential the highest occupied edge level lies at a different distance below it at every size, and a correction of relative order $1/L$ that depends on this distance does not extrapolate smoothly. We therefore averaged each sum over a Hann window of chemical potentials of half-width 0.08$\Delta_\mathrm{b}$, as an exact weighted sum over the occupations met in the window, and compared it with the prediction averaged over the same window. It was computed on the seven smallest sizes of each series, with up to 5184 sites. One control requires that, in the cases that the sums at fixed chemical potential resolve on those sizes, the level-averaged value lie at the same distance from its prediction, to within the larger of 0.01 and the spread of the sums at fixed chemical potential. The window holds 4 to 19 edge levels. The control holds in six of the seven cases, to within 0.0043, and fails in the seventh, the rectangle with the long side along $y$ at $\mu=\pm0.3877\Delta_\mathrm{b}$, by 0.049, where the level-averaged extrapolation is itself unresolved with a spread of 0.094. Because this control fails we draw no conclusion from this estimator. On all nine sizes the sum at fixed chemical potential of that seventh case is itself unresolved, at 2.3643 with a spread of 0.066, where the level-averaged sum on seven sizes gives 2.3637, so the control fails against a reference that does not hold. For the record, its spread is below 0.05 in the other eleven cases, among them the five that the sums at fixed chemical potential leave unresolved on the same sizes, and its extrapolations there lie within 0.0040 of the range of the prediction. Applied to the square at $t_y=0.6$ of the main series, with $L=24$ to 72, it gives 1.106--1.108 over the chemical potentials, against 1.104--1.106 from Eq.~\eqref{eq:kb} with the ribbon velocities. These numbers are a diagnostic and not a test.

\begin{table}[tp]
\caption{\label{tab:c2}Edge weight of the $|C|=2$ flakes. The columns give the shape, the chemical potential, the range of the prediction over the corner matchings (one number where its width is below $10^{-4}$), the extrapolated in-gap symmetric weight, the spread of its three extrapolation windows, and the extrapolated in-gap Hall magnitude. A spread above 0.05 marks an unresolved case.}
\begin{center}\small\begin{tabular}{lccccc}
\hline\hline
shape & $\mu/\Delta_\mathrm{b}$ & prediction & lattice & spread & Hall \\
\hline
square & $-0.1877$ & 2.1768--2.1847 & 2.1722 & 0.0149 & 2.0004 \\
square & $0.0123$ & 2.1753 & 2.1728 & 0.0145 & 1.9994 \\
square & $0.2123$ & 2.1773--2.1878 & 2.1693 & 0.0202 & 2.0005 \\
square & $\pm0.3877$ & 2.1904--2.2417 & 2.1226 & 0.1836 & 1.9999 \\
2:1, long side along $x$ & $-0.1877$ & 2.5012--2.5109 & 2.4798 & 0.0288 & 2.0002 \\
2:1, long side along $x$ & $0.0123$ & 2.4942 & 2.4986 & 0.0148 & 2.0003 \\
2:1, long side along $x$ & $0.2123$ & 2.5035--2.5163 & 2.5093 & 0.0196 & 1.9995 \\
2:1, long side along $x$ & $\pm0.3877$ & 2.5499--2.6140 & 2.6094 & 0.1575 & 2.0009 \\
2:1, long side along $y$ & $-0.1877$ & 2.3970--2.4059 & 2.4299 & 0.0804 & 2.0014 \\
2:1, long side along $y$ & $0.0123$ & 2.4005 & 2.4078 & 0.0256 & 2.0012 \\
2:1, long side along $y$ & $0.2123$ & 2.3959--2.4076 & 2.4108 & 0.0176 & 2.0013 \\
2:1, long side along $y$ & $\pm0.3877$ & 2.3801--2.4361 & 2.3643 & 0.0660 & 2.0002 \\
\hline\hline
\end{tabular}\end{center}
\end{table}

\section{Temperature and the circuit frequency}
\label{app:thermal}
The circuit frequency is set by the dipole response of the flake and the zero-point energy by its dipole fluctuation (Appendix~\ref{sec:coupling}). At temperature $T$ and fixed chemical potential, with $f_i$ the Fermi occupation of level $i$, a pair of levels $i$ and $j$ at energies $\varepsilon_i<\varepsilon_j$ enters the response with the factor $f_i-f_j$ and the fluctuation with the factor $f_i(1-f_j)+f_j(1-f_i)$. For a chiral band of constant velocity whose dipole matrix elements depend only on the level separation, the first factor sums to the same weight at every temperature and the second to the factor $\coth(\hbar\omega_n/2k_\mathrm{B}T)$ of a bosonic mode.
A lattice edge has a curved band and matrix elements that vary with energy, so the statement was tested on the ellipse with $L=64$ and the square with $L=48$ at $t_y=0.83$, $M=1.37$ and $\mu=0.0123\Delta_\mathrm{b}$. The thermal energy $k_\mathrm{B}T$ was set to 0.25, 0.5, 1, 2 and 4 times $\hbar\omega_\mathrm{EMP}$, and the sums run over all pairs of levels separated by less than $0.95\Delta_\mathrm{b}$. The response weight is the sum of $(f_i-f_j)s_{ij}$, with $s_{ij}$ the symmetric weight of the pair, its Hall weight is the same sum over the antisymmetric weights, the static polarizability weight divides each term by the level separation, and the fluctuation weight is the sum of $[f_i(1-f_j)+f_j(1-f_i)]s_{ij}$.
At zero temperature these sums reproduce the recorded in-gap weights to $4\times10^{-15}$. Up to $k_\mathrm{B}T=2\hbar\omega_\mathrm{EMP}$ the response weight, its Hall weight and the polarizability weight change by at most 0.0019, 0.0003 and 0.0016 of their zero-temperature values on the ellipse and by at most 0.0025, 0.0003 and 0.0110 on the square, inside the tolerance of 0.02. The fluctuation weight at $k_\mathrm{B}T=2\hbar\omega_\mathrm{EMP}$ is 4.06 and 3.96 times its zero-temperature value, and up to that temperature it stays within 0.011 of the bosonic form evaluated with the zero-temperature transitions.
At $k_\mathrm{B}T=4\hbar\omega_\mathrm{EMP}$ the thermal energy is 0.122 and $0.135\Delta_\mathrm{b}$, bulk carriers are no longer frozen out, and the response weight changes by 12 and 17\,\%, the polarizability weight by 18 and 30\,\%, and the Hall weight by 0.6 and 1.0\,\%. This breakdown was expected and lies outside the claim.

\paragraph*{The avoided crossing.}The splitting of the doublet was also computed directly. With the thermal response of the flake in place of $\chi$, the circuit of Appendix~\ref{sec:coupling} has the spectral function $\mathrm{Im}\,\{\omega_p^2-\Omega^2[1+\chi(\Omega)/C_pd^2]\}^{-1}$, in which each in-gap pair enters $\chi$ with the factor $f_i-f_j$ and a linewidth of 0.005 of the edge frequency. The circuit is tuned to the zero-temperature transition of largest weight along the field, and the coupling is set so that this transition alone would give a splitting of 0.12 or 0.28 of its frequency, the range of the device estimate. We required the doublet to stay resolved and its splitting to stay within 3\,\% of the zero-temperature value up to $k_\mathrm{B}T=2\hbar\omega_\mathrm{EMP}$ in all eight cases, two flakes with two field directions and two couplings each. The transition to which the circuit is tuned carries 0.92 to 0.99 of the in-gap weight along its axis. At zero temperature the largest deviation of the splitting from the value for that transition alone is 0.124\,\%. Up to $k_\mathrm{B}T=2\hbar\omega_\mathrm{EMP}$ it changes by at most 0.7\,\%, the dip between the two peaks stays below 0.005 of the smaller peak, and the peaks are at most 3.0 times as wide as those of a single mode. At $k_\mathrm{B}T=4\hbar\omega_\mathrm{EMP}$ the doublet is still resolved, its splitting has changed by up to 6.7\,\% and its peaks are about 7.3 times as wide at most. A single mode treated with the same code returns the exact splitting to $5\times10^{-5}$ of the mode frequency.
\end{appendix}

\bibliography{refs}

\end{document}